\documentclass[nofootinbib,floatfix,aps,12pt,amsmath,epsfig,groupedaddress,superscriptaddress]{revtex4} 
\usepackage{epsfig} 
\usepackage{graphicx} 
 
\begin{document}

\def\mymarginpar#1{\marginpar{\tiny\it\raggedright #1}} 
 
\newcommand{\aq}{$a_q$} 
\newcommand{\bq}{$b_q$} 
\newcommand{\cq}{$c_q$} 
\newcommand{\dq}{$d_q$}

\def\ra{\rangle} 
\def\la{\langle}

 
\begin{center} 
 
{\large
\bf FAST CARS: Engineering a Laser Spectroscopic Technique 
for Rapid Identification of Bacterial Spores} 
 
 
\thispagestyle{empty} 
\renewcommand{\thefootnote}{\fnsymbol{footnote}} 
 
{ 
M.~O.~Scully,\footnote{My friend and mentor Vicky Weisskopf used to say ``The
best way into a new problem is to bother people.'' This is faster than
searching the literature and more fun. I would like to thank my colleagues for
allowing me to be a bother and especially my coauthors who have suffered the most!
This paper is dedicated to Prof. Viktor von Weisskopf: premier physicist and
scientist--soldier who stood by his adopted  country in her hour of
need.}$^{1,2,3,5}$ 
G.~W.~Kattawar,$^{1,2}$ 
R.~P.~Lucht,$^{1,4}$ 
T.~Opatrn\'{y},$^{1,6}$ 
H.~Pilloff,$^{1}$ 
A.~Rebane,$^{7}$ 
A.~V.~Sokolov,$^{1,2}$ 
and 
M.~S.~Zubairy$^{1,2,8}$ }

 
\noindent {\it Institute for Quantum Studies$^{1}$, Dept. of 
Physics$^{2}$, Dept. of Electrical Engineering$^{3}$, 
Dept. of Mechanical Engineering$^{4}$, Texas A\&M 
University, College Station, Texas 
77843\\ 
$^{5}$Max-Planck-Institut f\"{u}r Quantenoptik, D-85748 Garching, 
Germany\\ 
$^{6}$Dept. of Theoretical Physics, Palack\'{y} University, 
Olomouc, Czech Republic\\ 
$^{7}$Dept. of Physics, Montana State University, Bozeman, 
Montana 59715, USA \\ 
$^{8}$Dept. of Electronics, Quaid-i-Azam University, 
Islamabad, Pakistan 
 
} 
 
(\today ) 
\end{center}

{\normalsize
\setlength{\baselineskip}{18pt}
Airborne contaminants, e.g., bacterial spores, are usually analyzed  by time
consuming microscopic, chemical and biological assays.  Current research into
real time laser spectroscopic detectors of  such contaminants is based on e.g.
resonance fluorescence.  The present approach derives from recent experiments
in which  atoms and molecules are prepared by one (or more) coherent  laser(s)
and probed by another set of lasers. These studies have  yielded such
counterintuitive results as lasers which operate  without inversion, ultra-slow
light with group velocities of  order 10 meters/sec, and generation of
ultra-short pulses of  light via phased molecular states. The preceding
examples are  based on inducing a phase coherent state of matter in the 
ensemble of simple molecules being studied. The connection with  previous
studies based on ``Coherent Anti-Stokes Raman  Spectroscopy" (CARS) is to be
noted. However generating and  utilizing maximally coherent oscillation in
macromolecules having  an enormous number of degrees of freedom is much more
challenging.  In particular, the short dephasing times and rapid internal
conversion  rates are major obstacles. However,  adiabatic fast  passage
techniques and the ability to generate combs of phase coherent  femtosecond
pulses, provide new tools for the generation and  utilization of maximal
quantum coherence in large molecules and  biopolymers. This extension of the
CARS technique is called FAST  CARS (Femtosecond Adaptive Spectroscopic
Techniques for Coherent  Anti-Stokes Raman Spectroscopy), and  the present
paper proposes and analyses ways in which it could be used to rapidly identify
pre-selected molecules in real time. 

}

\pagebreak 
 
 
\section{Introduction}

There is an urgent need for the 
rapid assay of chemical and biological unknowns, such as 
bioaerosols. Substantial 
progress toward this  goal has been made over the past decade. 
Techniques such as fluorescence  spectroscopy \cite{Cheng,Seaver99}, 
and UV resonant Raman spectroscopy 
\cite{Manoharan90,Nelson91,Ghiamati,Manoharan93} 
have been  successfully 
applied to the identification of biopolymers,  bacteria, and 
bioaerosols.

\begin{figure}[b] 
\centerline{\epsfig{file=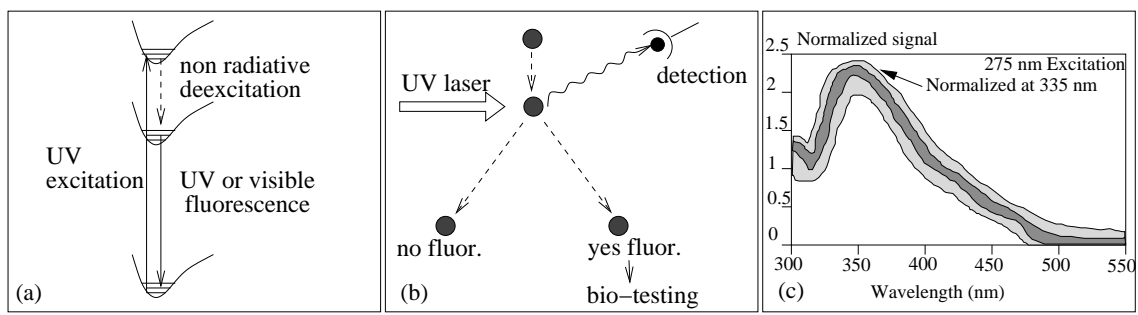,scale=0.4}} 
\parbox{90ex}{\caption{(a) Ultra-violet (UV) excitation radiation 
promotes molecules from ground state to an excited state 
manifold. This excited state manifold decays to the ground state 
via non-radiative processes to a lower manifold which then decays 
via visible or UV fluorescence. It is this fluorescence signature 
which is detected in resonance fluorescence spectroscopy; (b) 
Figure depicting a 
 scenario in which a UV laser interacts with dust particles and 
bio-spheres of interest. When, for example, a bacterial spore is 
irradiated, fluorescence will be emitted signaling that this 
particular system is to be further tested. In principle, uninteresting 
particles are deflected one way; but when 
fluorescence takes place, the particles are deflected in another 
direction and these particles are then subjected to further 
biological tests; (c) The shaded area displays the signal range for
the fluorescence spectrum of a number of biological samples, {\it 
Bacillus subtilis,\/} {\it Bacillus thuringiensis,\/} 
{\it Escherichia coli,\/} and {\it Staphylococcus aureus}. It 
is not  possible to distinguish between the different samples 
based on such a measurement (see \cite{Cheng} for more details). 
\label{Fig1}}} 
\end{figure}

At present field devices are being engineered \cite{Seaver99} which will 
involve an optical preselection stage based on, e.g., fluorescence  radiation
as in Fig. \ref{Fig1}.  If the fluorescence measurement  does not give the
proper signature  then that  particle is ignored. Most of the time the particle
will  be an uninteresting dust particle; however, when a signature  match is
recorded, then the particle is selected for special  biological assay, see Fig.
\ref{Fig1}b.  The  relatively simple fluorescence stage can very quickly sort
out  some of the uninteresting particles while the more time consuming 
bio-tests will only be used for the ``suspects''.

The good news about the resonance fluorescence technique is that 
it is fast and simple. The bad news is that while it can tell the 
difference between dust and bacterial spores, it can not 
differentiate between spores
and many other organic bioaerosols, see Fig. \ref{Fig1}c. 
 
However, in  spite of the encouraging success of the above mentioned studies, 
there is still interest in other approaches to, and tools for,  the rapid
identification of chemical and biological substances.  To quote from a recent
study \cite{Terror}: 
\begin{quote} 
``Current [fluorescence based] prototypes are a large improvement 
over earlier stand-off systems, but they cannot yet consistently 
identify specific organisms because of the similarity of their 
emission spectra. Advanced signal processing techniques may 
improve identification." 
\end{quote}

\begin{figure}[t] 
\centerline{\epsfig{file=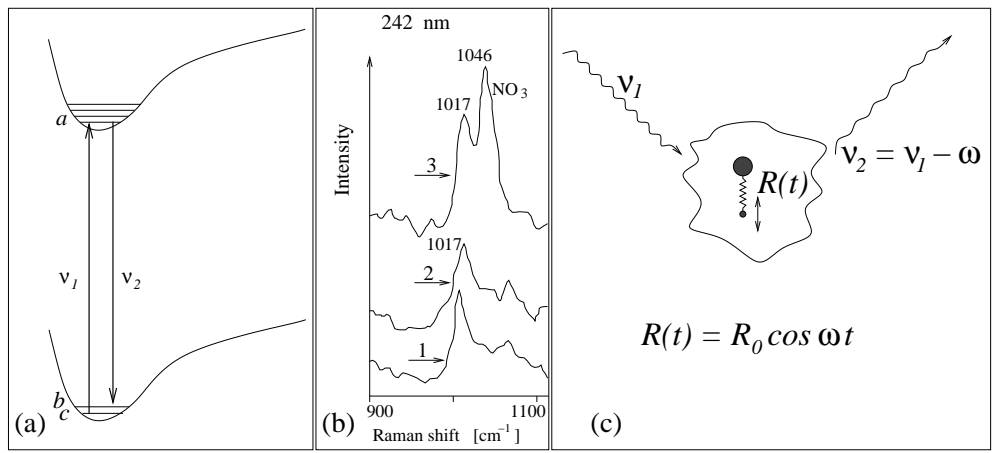,scale=0.25}} 
\parbox{85ex}{\caption{(a) Resonant Raman scattering  in which 
radiation $\nu_1$ excites the atom from $|c\ra$ to $|a\ra$ and 
the Stokes radiation is emitted taking the molecule from $|a\ra$ 
to $|b\ra$. The frequency of the excitation radiation is $\nu_1$, 
and the frequency of the Stokes 
radiation is $\nu_2$;   (b) 
Detail of UV resonance Raman spectra of spores of {\it Bacillus megaterium\/} 
(1),
{\it Bacillus cereus\/} (2), and of Calcium Dipicolinate (3), 
all excited at 242 nm; adapted from
W.H. Nelson and J.F. Sperry,
\textit{Modern Techniques in Rapid Microorganism Analysis,} 
edited by W.H. Nelson (VCH Publishers, N.Y. 1991).
\cite{Ghiamati} (see also Fig. \ref{FigNelson});
(c) Figure 
illustrating a more physical picture of Raman scattering in which 
a single diatomic molecule, consisting of a heavy nucleus e.g. 
Uranium 235 and a light atom e.g. hydrogen, scattering incident 
laser radiation at frequency $\nu_1$. The vibrational degrees of 
freedom associated with the diatomic molecule are depicted here 
as occurring with amplitude $R_0$ oscillating at frequency 
$\omega$. The scattered radiation from this vibrating molecule is 
at frequency $\nu_2 = \nu_1 - \omega$ for the Stokes 
radiation.  This classical picture of the vibrating 
dipole is to be understood as an amplitude times a sinusoidal 
oscillation at the frequency $\omega$ as indicated in the equation beneath 
the figure. $R_0$ is a quantum mechanically calculated 
oscillation amplitude as discussed in detail in Section III of 
the paper. \label{Fig02}}} 
\end{figure}

Resonant Raman spectra  hold promise for being spore specific as indicated in
Fig. \ref{Fig02}b.  This is the good news, the bad news is that the Raman 
signal is weak and it takes several minutes to collect the data  of Fig.
\ref{Fig02}b.  Since the through-put in a set-up such as 
that of Fig. \ref{Fig1}b  is large, the optical interrogation per particle  
must be essentially instantaneous. 

The question then is: Can we increase the resonant Raman signal  strength and
thereby reduce the interrogation time per  particle? If so, then the technique
may also be useful in various  detection scenarios.

The answer to the question of the proceeding paragraph is a 
qualified ``yes." We can enhance the Raman signal by increasing 
the coherent molecular oscillation amplitude $R_0$ indicated in 
\ref{Fig02}c. 
In essence this means maximizing the quantum coherence between 
vibrational states $|b\ra$ and $|c\ra$ of \ref{Fig02}a. 
 
Our point of view derives from research in the fields of laser  physics and
quantum optics which have concentrated on the  utilization and maximization of
quantum coherence. The essence of  these studies is the observation that an
ensemble of atoms or  molecules in a coherent superposition of states
represents, in a  real sense, a new state of matter aptly called ``phaseonium" 
\cite{ScullyZubairy}. 

In particular, we note that matter in thermodynamic equilibrium 
has no phase coherence between the electrons in the molecules 
making up the ensemble. This is discussed in detail in Section 
III. When a coherent superposition of quantum states is involved, 
things are very different and based 
on these observations, many interesting and counterintuitive 
notions are now a laboratory reality. These include lasing 
without inversion (LWI) \cite{LWI}, 
electromagnetically induced transparency (EIT) \cite{EIT}, 
light having ultra slow group velocities on the order of 10 
meters/sec \cite{slowlight}, 
and the generation of ultra short pulses of light based on phased 
molecular states \cite{sokolov}.

Another emerging technology central to the present paper is the  exciting
progress in the area of femtosecond quantum control of  molecular dynamics
originally suggested by Judson and Rabitz \cite{Rabitz}.  This is described and
reviewed in the articles by  Kosloff et al. \cite{Kosloff},  Warren, Rabitz and
Dahleh \cite{Warren93},  Gordon and Rice \cite{Gordon97},  Zare \cite{Zare}, 
Rabitz, de-Vivie-Riedle, Motzkus and Kompa \cite{Rabitz00},  and Brixner,
Damrauer and Gerber \cite{Brixner01}.  Other related work  on quantum coherent
control includes:  The quantum  interference approach of Brumer and Shapiro
\cite{Brumer86};  the  time-domain (pump-dump) technique proposed by Tannor,
Kosloff and  Rice \cite{Tannor86};  the stimulated Raman Adiabatic Passage
(STIRAP)  approach of Bergmann and co-workers \cite{Bergmann98} to  generate a
train of coherent laser pulses.  The preceding studies teach us how to  produce
pulses having arbitrary controllable amplitude and  frequency time dependence.
Indeed the ability to sculpt pulses by  the femtosecond pulse shaper provides
an important new tool for  all of optics, see the pioneering works by Heritage,
Weiner, and  Thurston \cite{Heritage},  Weiner, Heritage, and Kirschner
\cite{Weiner88}, Wefers and Nelson \cite{Wefers95} and Weiner \cite{Weiner00}.

An important aspect of the learning algorithm approach is that 
knowledge of the molecular potential energy surfaces and matrix 
elements between surfaces are not needed. Precise taxonomic marker frequencies
may not be known a priori; however, by using a pulse shaper coupled with a
feedback system, complex spectra can be revealed.

\begin{figure}[b] 
\centerline{\epsfig{file=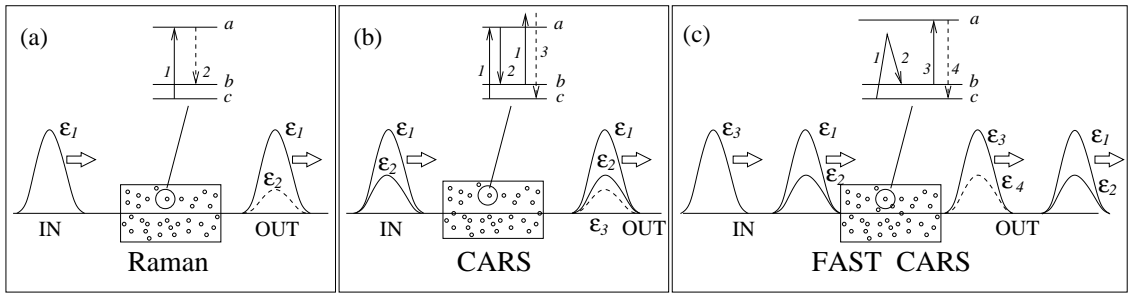,scale=0.3}} 
\begin{center} 
\parbox{90ex}{\caption{\label{Fig03}(a) Ordinary resonant Raman spectroscopy 
in which a drive laser of amplitude ${\cal E}_1$ generates a weak 
signal field having an amplitude ${\cal E}_2$. The incident 
signal consists of one pulse at $\nu_1$ and the pulse structure 
following interaction with the molecular medium consists of two 
pulses at $\nu_1$ and $\nu_2$. (b) The coherent Raman process 
associated with CARS is depicted in which two fields at frequency 
${\nu_1}$ and ${\nu_2}$ are incident with amplitudes ${\cal E}_1$ 
and ${\cal E}_2$. The third radiated anti-Stokes signal field at 
frequency $\nu_3$ is indicated. Hence CARS involves 2 fields in 
and 3 out. (c) FAST CARS configuration in which maximal coherent 
Raman spectroscopy is envisioned. The preparation pulses  ${\cal 
E}_1$ and ${\cal E}_2$ prepare maximum coherence between states 
$|b\ra$ and $|c\ra$. 
Next the probe laser ${\cal E}_3$ interacts 
with this oscillating molecular configuration and the anti-Stokes 
radiation is generated. Thus we have 3 fields in and 4 out when 
using FAST CARS. }} 
\end{center} 
\end{figure}

Thus, we now have techniques at hand for controlling trains of 
phase coherent femtosecond pulses so as to maximize molecular 
coherence. This allows us to increase the Raman signal while 
decreasing the undesirable fluorescence background. This has much 
in common with the CARS spectroscopy \cite{Demtroder} of Fig. \ref{Fig03}, 
but with essential differences as we now discuss. 
 
The presently envisioned improvement over ordinary CARS is based on enhancing
the  ground state molecular coherence. However, we note that molecules 
involving a large number of degrees of freedom will quickly  dissipate the
molecular coherence amongst these degrees of  freedom. This is a well known
difficulty and is addressed in the  present work from several perspectives.
First of all, when  working with ultra short pulses, we have the ability to
generate  the coherence on a time scale which is small compared with the 
molecular relaxation time. Furthermore, we are able to tailor the  pulse
sequence in such a way as to mitigate, and overcome  key limitations in the
application of conventional CARS to trace  contaminants. The key point is that
we are trying to induce  maximal ground state coherence, as opposed to the
usual situation within  conventional CARS  where the ground state coherence is
not a maximum as is shown  later in this paper. With FAST CARS (Femtosecond
Adaptive  Spectroscopic Techniques applied to Coherent Anti-Stokes  Raman
Spectroscopy) we can prepare the coherence between two  vibrational states of a
molecule with one set of laser pulses;  and use higher  frequency visible or
ultra-violet to probe  this coherence in a  coherent Raman configuration. This
will allow us to capitalize on  the fact that maximally coherent Raman
spectroscopy is orders of  magnitude more sensitive than incoherent Raman
spectroscopy. 

Having stated our goals and our approach toward attaining these goals,
we emphasize that the present paper represents essentially an engineering
endeavor. We propose to draw heavily on the ongoing work in quantum coherence
and quantum control as mentioned earlier.

For example, the careful experiments and analysis of the W\"{u}rzburg group on
the generation and probing of ground state coherence in porphyrin molecules
\cite{Heid01} by femtosecond-CARS (fs-CARS) are very germane to our
considerations. However, ground state coherence is not maximized in these
experiments.

In another set of beautiful experiments \cite{Chen00} they investigate the
selective excitation of polymers of diacetylene via fs-CARS. They control the
timing, phase and frequency (chirp) content of their preparation pulses. In
these experiments it was necessary to focus attention on the evolution of the
excited state molecular dynamics. We hope to avoid this complication as is
explained later.

Perhaps closest to our approach is the recent joint work of the Garching
Max-Planck and W\"{u}rzburg groups \cite{Zeidler02}. Their paper entitled
``Optimal control of ground-state dynamics in polymers'' is a prime example of a
FAST CARS experiment. However they concentrate on producing highly excited
states of the ``vibrational motion of a certain bond''. The application of their
technique to the production of maximum coherence between states $|b\rangle$ and
$|c\rangle$ of Fig. \ref{Fig02}a in a specific vibrational mode of their
molecule would be of great interest to us and is underway.

Finally we wish to draw the reader's attention to the useful collection of
articles in a recent special issue of the ``Journal of Raman Spectroscopy''
dedicated to fs-CARS \cite{JRS00}. Likewise the recent work of Silberberg and
coworkers \cite{Silberberg} in which they show that it is possible to excite one
of two nearby Raman levels, even when they are well within the broad fs pulse
spectrum is another excellent example of the power of the FAST CARS technique.

To summarize: the present work focuses on utilization of a  maximally phase
coherent ensemble of molecules, i.e. molecular  phaseonium, to enhance Raman
signatures. This will be accomplished  via the careful tailoring of a coherent
pulse designed to prepare the  molecule with maximal ground state coherence.
Such a pulse is a sort of  ``melody" designed to prepare a particular molecule.
Once we know  this molecular melody, we can use it to set that particular
molecule in  motion and  this oscillatory motion is then detected by another 
pulse; this is the FAST CARS protocol depicted in Figs \ref{Fig03}c  and
\ref{Fig13a}b.

In order to establish the viability and credibility of this  program, the
material covered in the present paper is presented  in  some detail. It is
hoped that scientists who are experts in  one phase of the subject, e.g.,
molecular biology but not with  subtleties of modern laser spectroscopy can
read the paper  without undue appeal to the literature or complicated
mathematical  developments. On the other hand, some basic facts of life,
endosporewise, are important. Hence, a short overview of some aspects of Raman 
spectroscopy as applied to macromolecules and especially to  biological spores
is presented. 

\renewcommand{\thefootnote}{\fnsymbol{footnote}}

In Section II, the status of Raman spectroscopy applied to  biological spores
is reviewed. In Section III, we compare various  types of Raman spectroscopy
with an eye to the recent successful  applications of quantum coherence in
laser physics and quantum  optics.  Section IV presents several experimental
schemes  for applying these considerations to  the rapid  identification of
macromolecules, in general, and biological  spores, in particular. Finally in
Section V we propose several scenarios in  which FAST CARS could be useful in
the rapid detection of  bacterial spores.   Where appropriate, mathematical 
details are included in Appendices and comparison between the  various types of
Raman spectroscopic techniques are discussed with  special emphasis on overall
sensitivity.  As stated earlier, the present paper is an engineering science
analysis of a promising approach to the problem of bacterial spore detection.
This is not a review paper. If the reader feels that we have missed or
misrepresented her research, we would be happy to learn how to use it to
improve our design and detection strategy. Indeed, we view this paper as
providing a point of departure, and will be pleased if it provokes discussion
and debate. If the reader is not provoked we apologize. It is very difficult to
annoy everybody in a single paper.

\pagebreak 

\section{Pico-Review of Raman spectroscopy applied to bacterial spores}
 
The bacterial spore is an amazing life form. Spores thousands of 
years old have been found to be viable. One textbook \cite{Black} 
reports that ``endospores trapped in amber for 25 million years 
germinate when placed in nutrient media." 
 
\begin{figure}[h] 
\centerline{\epsfig{file=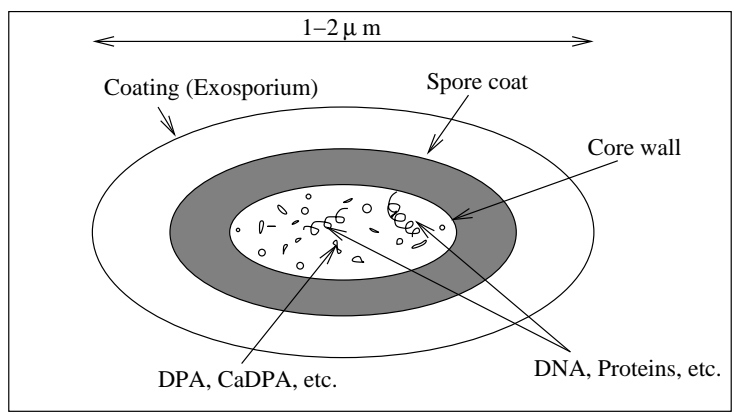,scale=0.5}} 
\parbox{90ex}{\caption{Sketch 
of spore indicating that the DPA and its salts, e.g., Ca-DPA are contained in the 
core and are in contact with the spore specific DNA ribosomes and 
cell proteins. 
\label{Fig4} } }
\end{figure}

A key to this incredible longevity is the presence of dipicolinic 
acid (DPA) and its salt calcium dipicolinate in the living core 
which contains the DNA, RNA, and protein as shown in Fig. 
\ref{Fig4}. 
 
A major role of the calcium DPA complex seems to be the removal 
of water, as per the following quote \cite{Talaro} 
: ``The exact 
role of these [DNA] chemicals is not yet clear. We know, for 
instance, that heat destroys cells by inactivating proteins and 
DNA and that this process requires a certain amount of water. 
Since the deposition of calcium dipicolinate in the spore removes 
water . . . it will be less vulnerable to heat."

\begin{figure}[h] 
\centerline{\epsfig{file=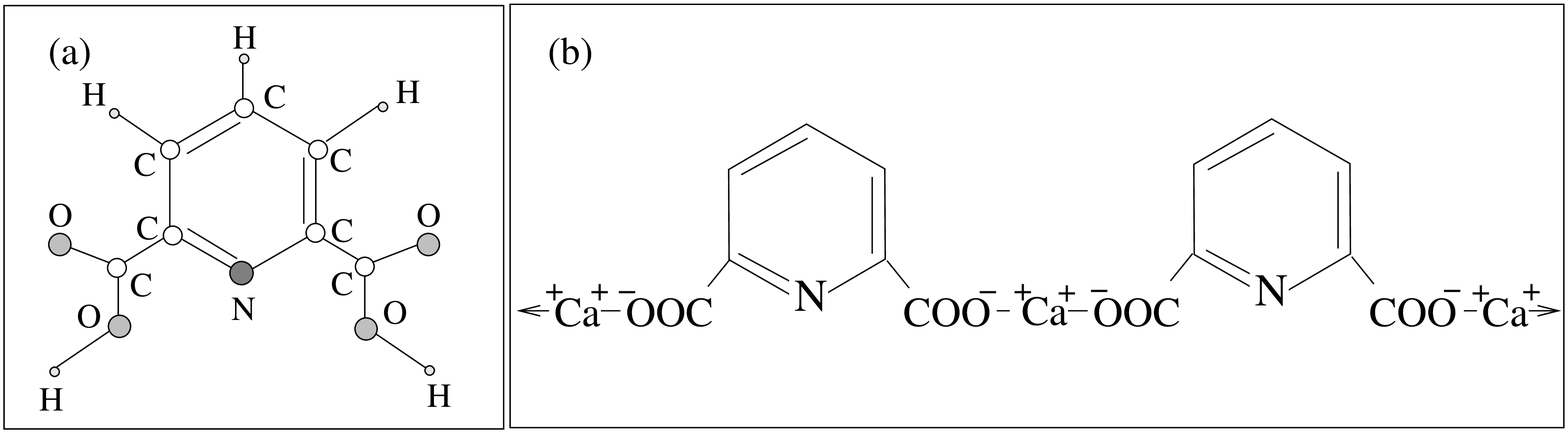,scale=0.4}} 
\parbox{90ex}{\caption{(a) 
Dipicolinic acid (2, 6-pyridinedicarboxylic acid, 
C$_5$H$_3$N(COOH)$_2$); (b) The Ca$^{2+}$ DPA complex. 
\label{FigDPA}}} 
\end{figure}

\begin{figure}[h] 
\centerline{\epsfig{file=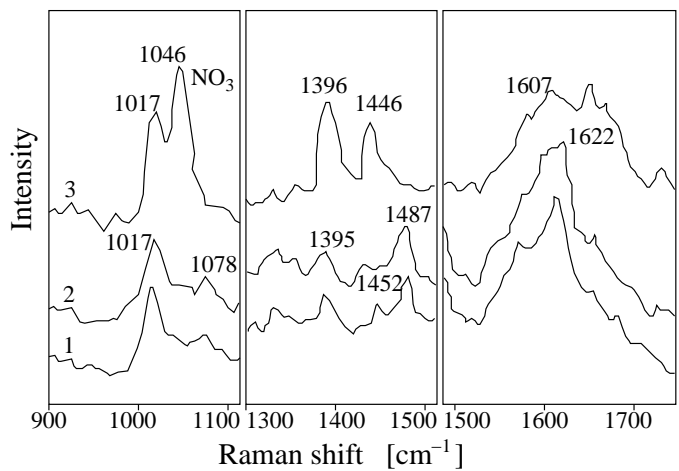,scale=0.5}} 
\parbox{90ex}{\caption{This figure 
(adapted from \cite{Nelson91})
shows  UV resonance Raman spectra of spores of
{\it Bacillus megaterium} 
(1), spores of {\it Bacillus cereus} (2), and 
calcium 
dipicolinate 
(3)
in three spectral regions. All samples are excited at 242 nm.
\label{FigNelson}} }
\end{figure}

Hence, one of the major components of bacterial spores is 
dipicolinic acid (DPA) and its ion as depicted in Fig. 
\ref{FigDPA}. Calcium dipicolinate can contribute up to $17\%$ of 
the dry weight of the spores.  A definitive demonstration 
\cite{Manoharan90} of 
this conjecture was made by comparing the 242 nm excitation 
spectra of calcium dipicolinate with spore suspensions of {\it 
Bacillus megaterium} and {\it Bacillus cereus}. {F}rom Fig. 
\ref{FigNelson}, 
it is seen that good matches were noted for the 
1017,  1396, 1446, and 1607 cm$^{-1}$ peaks of the calcium 
dipicolinate.

As has been emphasized by W. Nelson and coworkers 
\cite{Manoharan90,Nelson91,Ghiamati,Manoharan93}, the 
presence of DPA and its calcium salt gives us a ready made marker 
for endospores. As has been mentioned earlier and as will be 
further discussed later, this is the key to Raman fingerprinting 
of the spore. 
 
We note however that fluorescence spectroscopy was one of the 
first methods used for detection of bacterial taxonomic markers 
and is still used for detection where high specificity is not 
required. This technique is an important addition to the ``tool 
kit" of scientists and engineers working in this area.

A possible FAST CARS protocol is as follows: First we obtain size 
and fluorescence information. If this is consistent with the 
presence of a particular bacterial spore we could then 
automatically perform a FAST CARS analysis sensitive to DPA so as 
to further narrow the number of suspects. 
 
It is important to note that just as resonant Raman is some 
$10^6$ times more sensitive than non-resonant, coherent Raman 
yields a much stronger signal than ordinary incoherent Raman 
spectroscopy. This makes it possible to collect the Raman spectra 
much more rapidly via FAST CARS and this is very important in the 
ultimate scheme of things. 
 
To summarize: we will generate quantum coherence in 
macromolecules by working with the now available femtosecond 
pulse trains in which there exists phase coherence between the 
individual pulses. In this way, one can enhance coherent Raman 
signatures. The utilization of ``molecular music" to generate 
maximal  phase coherence holds promise for the identification and 
characterization of macro and bio-molecules.

\pagebreak 
 
\section{Comparison of different types of Raman spectroscopy} 
 
Raman scattering is an inelastic scattering of electromagnetic  fields off
vibrating molecules. The origin of Raman scattering  dates back to a
theoretical paper in Naturwissenschaften by A.  Smekal in 1923 entitled
(translated) ``The quantum theory of  dispersion'' \cite{Smekal}.  It was
followed by another paper in a 1923 Physical Review (by A.~Compton) entitled
``A quantum theory of the scattering of X-rays  by light elements''
\cite{Compton}.  Some historians feel that these two papers gave C.V. Raman
the  idea for the experiments that were performed with K.S. Krishnan  and led
to the discovery of the effect in over 60 liquids. Raman  and Krishnan
published their results entitled ``A new type of  secondary radiation'' in
Nature on March 28, 1928  \cite{Raman}.  It was soon followed by the landmark
paper of G. Landsberg and L.  Mandelstam who found the same effect in quartz
and published a  paper entitled (translated) ``A novel effect of light
scattering  in crystals'' which appeared on July 13, 1928 in 
Naturwissenschaften \cite{Landsberg}.  By the end of 1928  dozens of papers had
already been published on the ``Raman''  effect. 

In this section we first recall the quantum mechanical picture of a vibrating 
molecule. We then discuss the principles of different types of Raman 
spectroscopy. 
 
\subsection{Molecular vibrations} 
 
\begin{figure}[h] 
\centerline{\epsfig{file=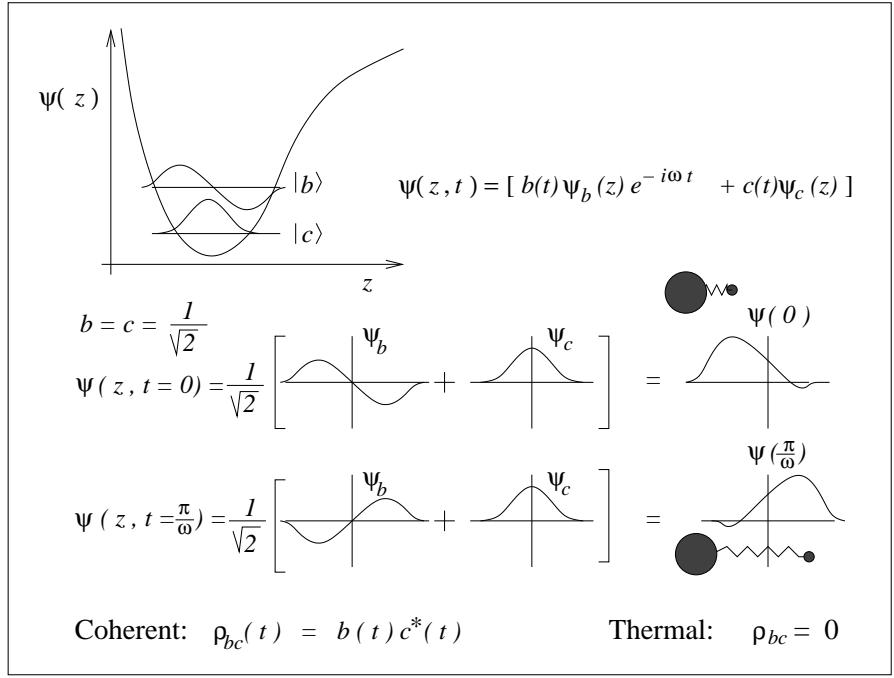,scale=0.43}} 
\parbox{90ex}{\caption{Molecular coherence: atomic distance of a 
molecule in a coherent superposition of the 
vibrational states $|b\rangle$ and $|c\rangle$ oscillates periodically with 
time. On the other hand, a molecule in a 
thermal equilibrium state is stationary and its 
coherence $\varrho_{bc}$ vanishes. 
\label{Fig2}} }
\end{figure}

Let us consider a simple diatomic molecule for 
explanation of the principle. The interatomic oscillation
can be visualized via
a classical picture of the vibrating molecule as 
in Fig. \ref{Fig02}c. Quantum mechanically, the situation can be understood as 
depicted in Fig. \ref{Fig2}. The potential energy of the molecule depends on 
the interatomic distance $z$ and has a well pronounced minimum. The Hamiltonian 
of the vibrating molecule has a set of discrete eigenstates; in Fig. \ref{Fig2} 
we show just the ground state $|c\rangle$ and the first excited state 
$|b\rangle$. Whereas in each of these states the mean displacement from the 
equilibrium position is zero, a quantum {\em superposition\/} of these states 
has generally a nonzero mean displacement $R(t)$ which varies with time. 
Assuming that in time $t=0$ the molecule is in a superposition state 
$|\Psi(0)\rangle = b|b\rangle + c|c\rangle$, then in time $t>0$ 
the state is $|\Psi(t)\rangle = 
b|b\rangle + c\exp (-i\omega t)|c\rangle$, where the frequency 
$\omega = (E_c-E_b)/\hbar$ is the difference of the energies of the eigenstates 
$|c\rangle$ and $|b\rangle$ divided by the Planck 
constant $\hbar$. The mean 
displacement $R(t)$ is then $R(t) = \langle \Psi(t)| \hat R |\Psi(t) \rangle$ is 
then 
\begin{eqnarray} 
 R(t) = R_0 \exp (\omega t - \varphi_0) + \text{c.c.}, 
\end{eqnarray} 
where 
\begin{eqnarray} 
 R_0 =  \left| b c^{*} \right| \langle b | \hat R | c \rangle  
 \label{r0bc} 
\end{eqnarray} 
is the displacement amplitude, $\varphi_0$ is the initial phase 
(determined by the phases of the coefficients $b$ and $c$), and 
$\hat R$ is the displacement operator. It can be seen from Eq. 
(\ref{r0bc}) that one can reach the maximum amplitude of the mean 
displacement if the superposition coefficients $b$ and $c$ are of 
the same magnitude, i.e., $|b| = |c| = 1/\sqrt{2}$ so that 
$R_0^{\text{max}}=\langle b|\hat R |c\rangle /2$. Thus the product 
$bc^*$ is of special importance for determining the vibrational 
amplitude.
This coherent superposition of states is generally described by 
the {\em off-diagonal density matrix element}  $\rho_{bc}$
which for the present simple case is given by $\rho_{bc} = bc^*$.
Without going into  detail we simply state that the 
density matrix element $\rho_{bc}$ is a complex number ($0 \leq 
|\rho_{bc}| \leq 1/2$) characterizing the quantum state of the 
molecule and determining the amplitude of the mean displacement. 
For some quantum states quantum coherence is not present (e.g., 
energy eigenstates, thermal states, etc.), whereas for some 
states it can reach the maximum magnitude (i.e., for 
$|\Psi\rangle = 2^{-1/2}(|b\rangle + |c\rangle)$).

We emphasize that when all molecules are in the same 
superposition state, the response of the sample to an optical 
signal is very different from the  thermal state. 
Preparation and optical probing of molecular vibrations
is the essence of Raman spectroscopy.

 
\subsection{Classical description of the Raman scattering} 
 
The simplest classical description of the Raman effect assumes that the 
polarizability $\alpha$ of the molecule is dependent on the 
relative positions of the 
atomic nuclei. The polarizability is the proportionality factor between the 
external electric field $E$ and the molecular dipole moment $P$, 
$P = \alpha E$, 
and for a vibrating molecule it is a time-dependent quantity. In the linear 
approximation, and assuming just the scalar case, 
the polarizability can be written as 
$\alpha (Q) =  \alpha (0) +  \alpha ' Q$, 
where $Q$ is the generalized coordinate of the vibrating molecule, $\alpha (0)$ 
is the polarizability of the equilibrium state, and $\alpha ' = \partial \alpha 
/ \partial Q$ with $Q=0$. If the molecule vibrates with frequency $\omega_{bc}$, 
the coordinate $Q$ changes as $Q = Q_{0} \cos (\omega_{ab} t)$. The electric 
field irradiating the molecule oscillates as $E = E_{0} \cos (\nu_{1} t)$. 
Thus, one finds that the 
molecular dipole oscillates with several frequencies: 
with the frequency of the incoming radiation 
$\nu_{1}$ (leading to the Rayleigh 
scattering), and with the shifted frequencies 
$\nu_{1}\pm \omega_{bc}$ (leading to the Stokes and anti-Stokes 
Raman frequencies). 
 
Even though this model is able to predict the correct 
frequencies of the scattered light, it cannot tell us anything about the 
intensities of different field components
in spontaneous scattering processes. To get more information about the 
scattering process, one needs a quantum mechanical model of the molecule. 
Let us now study the main features of the various Raman scattering processes.

 
\subsection{Stokes vs. anti-Stokes scattering} 
 
\begin{figure}[h] 
\centerline{\epsfig{file=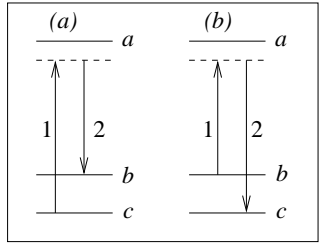,scale=0.4}} 
\begin{center} 
\parbox{90ex}{\caption{Stokes 
(a) and anti-Stokes (b) Raman scattering. Pump field 1 
interacts with a vibrating molecule to produce the scattered 
field 2 which has either lower frequency (Stokes scattering) or 
higher frequency (anti-Stokes scattering). \label{Fig6}} }
\end{center} 
\end{figure}

Raman scattering is an optical phenomenon in which there is a 
change of frequency of the incident light. Light with frequency 
$\nu_1$ scatters inelastically off the vibrating molecules such 
that the scattered field has frequency $\nu_2 = \nu_1 \pm 
\omega_{bc}$, where $\omega_{bc}$ is the frequency of the 
molecular vibrations. The field with down-shifted frequency 
$\nu_2 = \nu_1 - \omega_{bc}$ is called Stokes field and its 
generation corresponds to the process depicted in Fig. \ref{Fig6}a, whereas 
the frequency up-shifted radiation is called  the anti-Stokes 
field and corresponds to the process in Fig. \ref{Fig6}b. 
 
\subsection{Spontaneous vs. stimulated Raman scattering} 
 
There are two basic Raman processes: the so-called spontaneous and 
stimulated Raman scattering. Spontaneous scattering occurs 
if a single laser beam with intensity below a certain 
threshold illuminates the sample. In condensed matter, in 
propagating through 1 cm of the scattering medium, only 
approximately 10$^{-6}$ of the incident radiation is typically 
scattered into the Stokes field (see, e.g., \cite{Boyd}). 
Stimulated scattering which occurs with a very intense 
illuminating beam is a much stronger process in which several 
percent of the incident laser beam can be converted into the 
other frequencies. From the quantum-optical point of view, the 
Raman scattering can be described by means of photon numbers 
occupied in different modes. The rate of photon number increase 
in the Stokes mode can be written as 
$ \dot n_{S} = \eta n_{L}(n_{S}+1),$ 
where $n_{L}$ is the number of photons in the incident laser mode 
and $n_{S}$ is the number of photons in the Stokes mode. Here 
$\eta $ is a proportionality constant. 
 
For spontaneous Stokes scattering $n_{S} \ll 
1$, and the intensity of the scattered field is roughly 
proportional to the length traveled by the incident field in the 
medium. On the other hand, for $n_{S}>1$, the stimulated process 
becomes dominant and the scattered field intensity can increase 
exponentially with the medium length.

 
\begin{table}[b] 
\begin{tabular}{|@{\qquad}l|r @{\quad}|r @{\quad}|} 
\hline {\bf Process}  & ~ {\bf Raman Coherence} $\varrho_{cb}$ ~ & 
{\bf Dipole  Coherence} $\varrho_{ab} ~ \ ~ $ 
\\ \hline \hline 
\begin{tabular}{c} 
 Raman  \\  {\epsfig{file=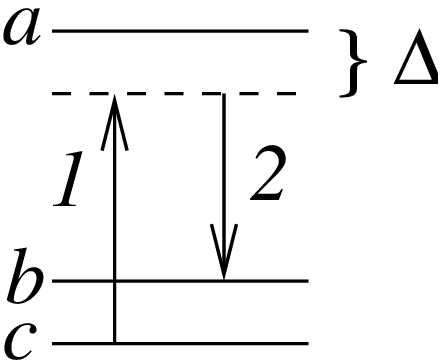,scale=0.4}} \\ (Weak drive) 
\end{tabular} 
 & 
\begin{tabular}{lr} 
 ~ & ~ \\
 $\displaystyle i \frac{\Omega_2 \Omega_1^*}{\gamma_{bc}\Delta}$ 
 ~ \qquad \ ~ & ~  \\
 ~ & $10^{-5}$  
\end{tabular} 
 & 
\begin{tabular}{lr}
 ~ &  ~ \qquad (incoh.) \\
 $\displaystyle -\frac{\Omega_2}{\Delta} 
 \frac{|\Omega_1|^2}{\Delta \gamma_{bc}}$ ~ \ ~ & ~ \\
 ~ & $10^{-9}$  
\end{tabular} 
\\ 
\hline 
\begin{tabular}{c} 
Resonant Raman  \\   {\epsfig{file=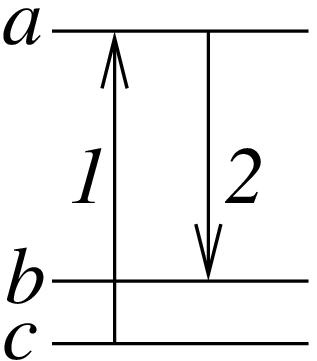,scale=0.4}} \\ (Weak drive) 
\end{tabular} 
 & 
\begin{tabular}{lr}
 ~ & ~ \\
 $\displaystyle - \frac{\Omega_2 \Omega_1^{*}}{\gamma_{ac}\gamma_{bc}}$ 
 ~ \quad ~
 & ~ \\
 ~ & ~ \quad $10^{-2}$ 
\end{tabular} 
  & 
\begin{tabular}{lr}  
 ~ &  ~ \qquad (incoh.) \\
 $\displaystyle -\frac{\Omega_2}{\gamma_{ab}} 
 \frac{|\Omega_1|^2}{\gamma_{ac} \gamma_{bc}}$ ~ \ ~ & ~ \\
 ~ & $10^{-3}$ 
\end{tabular} 
\\
\hline 
\begin{tabular}{c} 
Raman   \\  {\epsfig{file=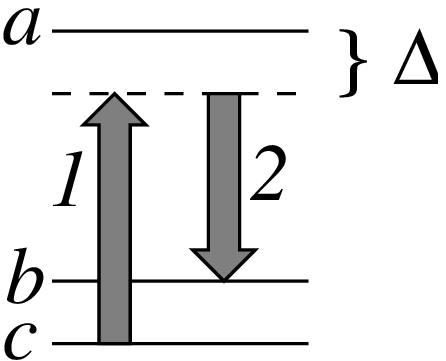,scale=0.4}}  \\ (Strong drive) 
\end{tabular} 
 & 
\begin{tabular}{lr} 
~ & (max. coh.) \\
 $\displaystyle \frac{i}{4} \sqrt{\frac{\gamma_{1}}{\gamma_{bc}}}$ \ ~ &
 ~ \\
 ~ & $10^{-3}$
\end{tabular} 
 & 
\begin{tabular}{lr} 
 ~ & ~ \qquad  (max. coh.) \\
 ~ \quad $\displaystyle i \frac{1}{4} \frac{\Omega_2}{\Delta} 
 \sqrt{\frac{\gamma_{1}}{\gamma_{bc}}}$ & ~ \\
 ~ & $10^{-6}$ 
\end{tabular}
\\ 
\hline 
\begin{tabular}{c} 
Resonant Raman  \\   {\epsfig{file=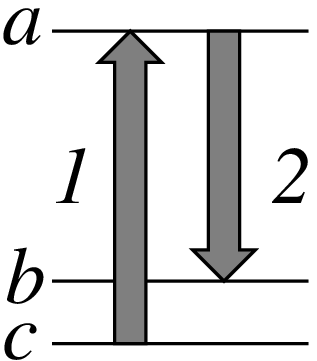,scale=0.4}} \\ (Strong drive) 
\end{tabular} 
 & 
\begin{tabular}{lr}
 ~ & ~ \qquad (max. coh.) \\ 
 $\displaystyle \frac{1}{2}$  & ~ \\
 ~ & $10^0$
\end{tabular}
& 
\begin{tabular}{lr}
 ~ &  ~ \qquad (max. coh.) \\
 $\displaystyle  \frac{i \Omega_1}{2\gamma_{ab}}$  & ~ \\
 ~ & $10^{-1}$
\end{tabular} 
\\ 
\hline 
\end{tabular}  
\caption{Comparison of different Raman spectroscopic techniques
as derived in Appendix~\ref{App-Dream}. 
The density matrix element $\rho_{bc}$ governs the amplitude of 
coherent vibration,
whereas the element $\rho_{ab}$ is  
proportional to the electronic polarization responsible for 
emission of radiation.
$\Omega_{1,2}$ are the Rabi frequencies,  
$\Delta$ is the detuning of the electronic transition,  
$\gamma_{ab}$, $\gamma_{ac}$ are the decay rates of the optical transitions,  
$\gamma_{bc}$ is the decoherence rate of the  vibrational states, and
$\gamma_1$ is the
decay rate  from level $b$ to $c$. 
The approximated values (shown in the lower right corner)
were obtained for $\gamma_{ab} \approx \gamma_{ac}
\approx \gamma_{bc} \approx 10^{12}$s$^{-1}$,  $\gamma_{1}
\approx 10^6$s$^{-1}$, $\Delta \approx 10^{15}$s$^{-1}$, and 
$\Omega_{1,2} \approx 10^{11}$s$^{-1}$ for weak driving and
$\Omega_{1,2} \approx 10^{12}$s$^{-1}$ for strong driving.
Note that $\Omega \approx 10^{11}$s$^{-1}$ corresponds to a 10~ns pulse
with 0.1~mJ energy focused on a square millimeter spot if the electronic
transition dipole moment is $\wp \approx 10^{-19}$C$\times 10^{-10}$m
[see Eq. (\ref{Omega-Rabi})].
}
\label{Tab1} 
\end{table} 

\subsection{Resonant vs. non-resonant Raman processes} 
 
The resonant Raman process (appearing when 
the frequency of the incident radiation coincides with one of the
electronic transitions)
is much richer than the nonresonant, 
and we now turn to a discussion of the resonant problem. 
 
Resonant Raman radiation is governed by the oscillating dipole 
between states $|a\ra$ and $|b\ra$ (Stokes) and/or $|a\ra$ and 
$|c\ra$ (anti-Stokes) in the notation of Fig. \ref{Fig03} and  
Table \ref{Tab1}. In the Stokes case, the steady state coherent oscillating 
dipole $P(t)$, divided by the dipole matrix element 
$\wp_{ab}=e\la a|r|b\ra$, is the important quantity. That is 
$\rho_{ab}(t)\equiv P(t)/\wp_{ab}$, as given by Eq. (\ref{raman-rho-ab}), 
is 
\begin{eqnarray} 
\varrho_{ab}=-i\left[\Omega_2(n_a-n_b)-
\Omega_1\varrho_{cb}\right]/\left[\gamma_{ab}-i(\omega_{ab}-\nu_2)\right]
\label{eq-res-raman} 
\end{eqnarray} 
where the Raman coherence is $\rho_{bc}$ as discussed earlier, 
e.g. Fig. \ref{Fig2}. In Eq. (\ref{eq-res-raman}), $\omega_{ab}$ is the
transition frequency between the electronic states $a$ and $b$, $\nu_2$ is the
frequency of the generated field, and the other 
quantities  are 
defined in the caption of Table \ref{Tab1}. 

The main advantage of resonant Raman scattering is that the signal is very
strong---up to a million times stronger compared to the signal of 
nonresonant scattering \cite{Nelson91}.  It is also very useful that only those
Raman lines corresponding to very few vibrational modes associated with
strongly absorbing locations of a molecule show this huge intensity
enhancement. On the other hand, the resonance Raman spectra may be contaminated
with fluorescence. However, this problem can be avoided by using UV light so
that most of the fluorescence appears at much longer wavelengths than the Raman
scattered light and is easily filtered out.


\subsection{Coherent vs. incoherent Raman scattering from many molecules} 
 
\begin{figure}[h] 
\centerline{\epsfig{file=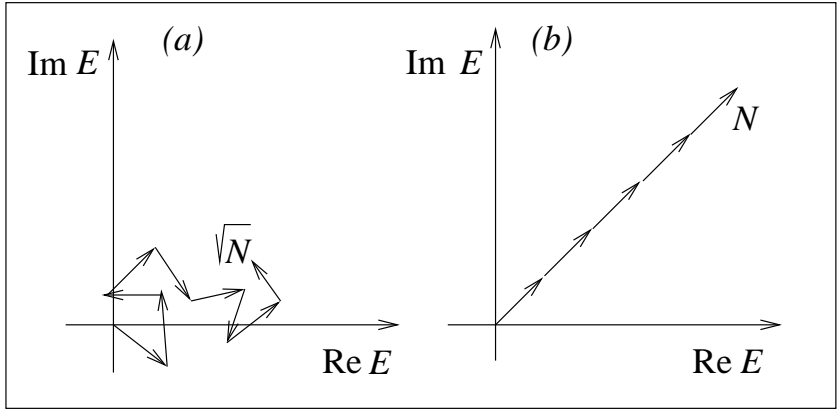,scale=0.5}} 
\begin{center} 
\parbox{90ex}{\caption{Incoherent 
$(a)$ vs. coherent $(b)$ addition of the radiation 
from $N$ molecules. Each arrow represents the contribution of one molecule
to the resulting electric field vector.
\label{Fig8}} }
\end{center} 
\end{figure}

An important distinction between different Raman scattering 
schemes is based on the phase relation of the field scattered off 
different molecules. In the incoherent case  the  spontaneous 
contributions of individual molecules sum up with random phases. 
The magnitude of the emitted electric field then scales as 
$\sqrt{N}$ (see Fig. \ref{Fig8}a). On the other hand, if all the molecules 
are prepared in the same coherent superposition of their states 
$|b\rangle$ and $|c\rangle$, their contributions to the emitted 
field have the same phases and the magnitude of their sum is 
proportional to $N$ (see Fig. \ref{Fig8}b). Thus, the intensity radiated 
from the sample is proportional to $N$ in the incoherent case and 
to $N^2$ in the coherent case. 
 
\begin{figure}[b] 
\centerline{\epsfig{file=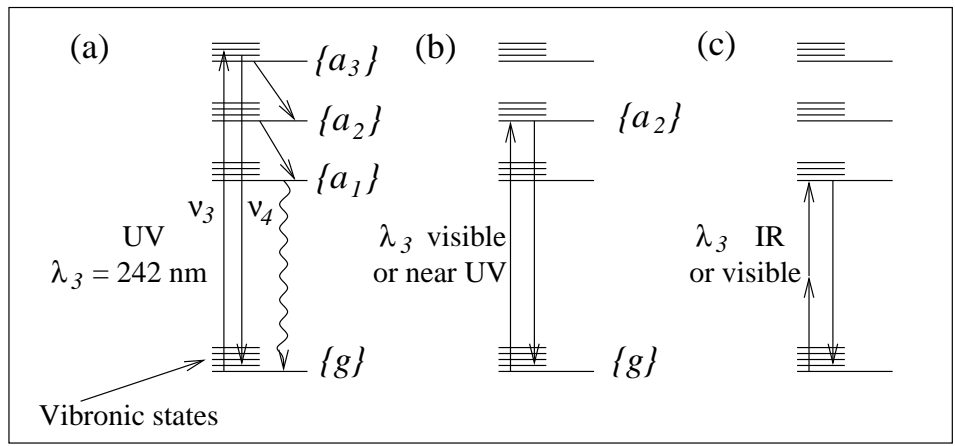,scale=0.5}} 
\parbox{90ex}{\caption{(a) Figure depicting resonant Raman 
experiments at 242 nm in which the molecule is excited to a high 
lying electronic state while fluorescence occurs at longer 
wavelengths. (b) By utilizing molecular coherence it is possible 
to enhance the rate of production of Stokes signal and mitigate 
fluorescence. (c) Scattering of the high frequency field at 
$\lambda_3$ off surfaces and other states can also be a source of 
confusion and this is eliminated by using a two photon IR drive. 
\label{Fig7}} }
\end{figure}

There is another important aspect of the coherent vs. incoherent 
resonant scattering process, namely the rate of emission from 
the $N$ atom ensemble. In the far off resonance, Smekel-Raman limit, 
the emission and absorption are simultaneous since the 
transitions to the excited state(s) are virtual. This is not 
true for the resonant Raman processes. In that case the molecule is 
excited to the $|a\ra$ state of Fig. \ref{Fig7} where, in the case of 
weak Stokes field $\varepsilon_c$, it can live for many 
nanoseconds. 
 
However, the cooperative emission rate from $N$ molecules can be 
much faster than that of a single molecule. As was shown by 
Dicke, and derived in Appendix \ref{App-Dicke}, if the spontaneous emission 
lifetime of $N$ incoherent dipoles is given by $\tau$; then the 
cooperative spontaneous emission lifetime of $N$ coherently 
prepared dipoles can be as short as $\tau/N$. 
 
This superradiant ``speed up" of the radiation process can have 
important consequences for the present problem. We recall 
that the experiments of Nelson and coworkers 
\cite{Manoharan90,Nelson91,Ghiamati,Manoharan93}
are carried out at 
242 nm so as to ``ride above" the fluorescence noise, see Fig. 
\ref{Fig7}b. But, if we can enhance the spontaneous emission rate so 
that the transition rate from the ${a_2}$ manifold to the ground 
state $g$ is faster than from the internal (non radiative) rate 
from ${a_2}\rightarrow {a_1}$, it would be possible to 
mitigate fluorescence noise. Then it would be possible to carry 
out resonant Raman studies with visible or near UV lasers instead 
of using the 242 nm wavelength.

 
\pagebreak 
\section{FAST CARS} 
 
\subsection{Generation of atomic coherence} 
\label{Subsec-Generation} 
The purpose of this section is to demonstrate the utility of pulse 
shaping as a mechanism for generating maximal coherence. The 
Raman signal is optimized at the condition of maximal molecular 
coherence.  When in this state, 
each of the molecules oscillates at a maximal amplitude, and all 
molecules in an ensemble oscillate in unison. Here we discuss 
several methods for the preparation of maximal coherence state. 
 
\subsubsection{Adiabatic Rapid Passage via Chirped Pulses} 
 
\begin{figure}[h] 
\begin{center} 
\includegraphics[scale=0.5]{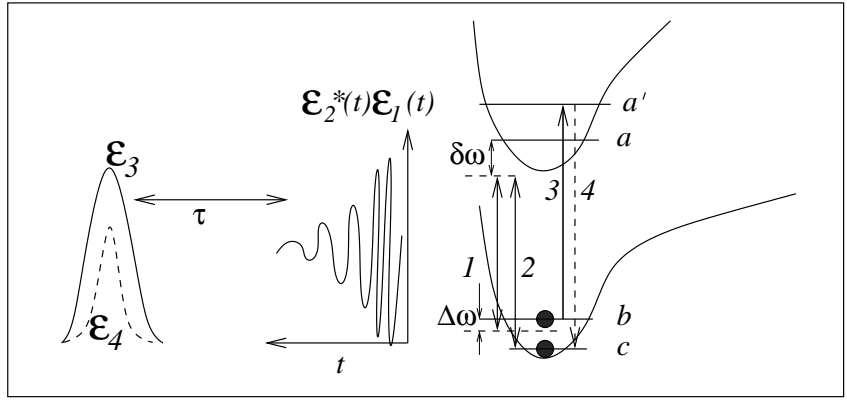}
\end{center} 
\parbox{90ex}{\caption{\label{figprepcoher2} 
Energy level schematics for a three-level 
system to generate maximum coherence between the levels 
$|b\rangle$ and $|c\rangle$ via fields ${\cal E}_1$ and ${\cal E}_2$. 
These fields are off-resonant with the 
electronic detuning $\delta \omega$ and possibly also with
the Raman detuning $\Delta \omega$ which can vary in time,
thus chirping the pulses. 
After preparing the coherence $\rho_{bc}$ with fields
${\cal E}_{1,2}$, the probe field ${\cal E}_{3}$ 
gives rise to the anti-Stokes field ${\cal E}_4$.}}
\end{figure}

A particularly simple and robust approach to the generation of 
the maximal coherence is to use a detuning $\delta \omega$ which is 
largely independent of inhomogeneous broadening and variations in matrix
elements (Fig. \ref{figprepcoher2}). 
 
Such multilevel molecular system can be described in terms  of an effective
two-by-two Hamiltonian \cite{CombGeneration}.  Diagonalization of this
Hamiltonian (Appendix \ref{App-Eigenstates})  allows us  to analyze the
evolution of the system by drawing analogies to  2-state systems. If the
excitation is applied resonantly  ($\Delta\omega=0$), such that the initial
state of the system  (the ground state $\vert c \rangle$ is projected onto the
new  basis formed by the eigenvectors $\vert + \rangle$ and $\vert -  \rangle$
[Eq. (\ref{eqC2})], the system undergoes a sinusoidal Rabi
flopping between  states $\vert b \rangle$ and $\vert c \rangle$.  In this 
situation one can choose to apply a $\pi/2$ pulse in order to  create the
maximal coherence $\vert \rho_{bc} \vert = 0.5$.

Alternatively, one can apply an excitation at a finite detuning 
$\Delta\omega$, to allow all population, which is initially in 
the ground state, to follow the eigenstate $\vert + \rangle$ 
adiabatically.  The coherence $\rho_{bc}$ is then 
\begin{equation} 
\rho_{bc} =  \frac{1}{2} \sin \theta \ e^{i\varphi} \eqnum{7} 
\end{equation} 
 
For molecular systems with large detunings, the Stark shifts $A$  and $D$ are
approximately equal and $\theta \cong \tan^{-1}  \left(2 \vert B \vert/2
\Delta\omega \right)$, where $B$ is the effective Raman Rabi frequency
(Appendix \ref{App-Eigenstates}). One method of  achieving the condition
$\vert\rho_{bc}\vert=0.5$ is to choose  $\Delta\omega$ and to increase the
product of the two incident  fields until $\theta$ is near $90^{\circ}$. This
is done  adiabatically with the product of the fields changing slowly as 
compared to the separation of the eigenvalues. Instead, at a  fixed field, one
may allow $\Delta\omega$ to chirp from an  initial value toward zero. 

We note that earlier, Grischkowsky \cite{Grischkowsky} and Oreg 
et al. \cite{Oreg} have described preparation mechanisms in 
two-state and multi-state systems, and Kaplan et al. 
\cite{Kaplan} have predicted existence of $2\pi$ Raman solitons. 
 
\subsubsection{Fractional STIRAP} 
 
\begin{figure}[h] 
\begin{center} 
\includegraphics[scale=0.4]{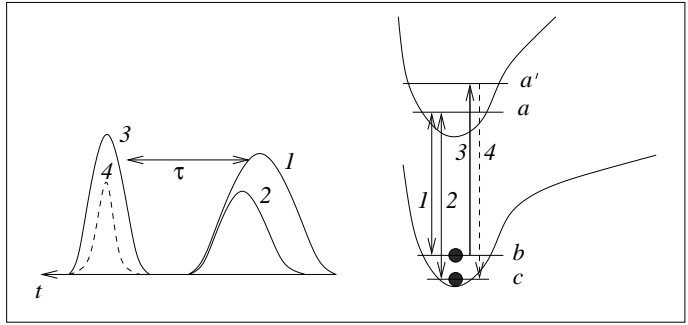} 
\end{center} 
\parbox{90ex}{\caption{Energy 
level schematics for the generation of maximum coherence 
between the levels $|b\rangle $ and $|c\rangle $ via fractional 
STIRAP by counterintuitive pulses 1 and 2. After a time delay of 
$\tau $ the pulse resonant with $|{a'} \rangle\rightarrow 
|b\rangle$ transition produces a signal at $\omega_{{a'}c}$. 
\label{Fig9}} }
\end{figure}

In an all-resonant $\Lambda$ scheme (Fig. \ref{Fig9}, with $\delta 
\omega = \Delta \omega =0$) maximal coherence can be prepared 
between the levels b and c in a fractional 
stimulated Raman adiabatic passage (STIRAP) set up by a 
counterintuitive pulse sequence \cite{Bergmann98,vss, Jain}, such that the 
population of the upper state a is always zero and 
fluorescence from this state is eliminated.  This can be 
accomplished via a counterintuitive sequence of two pulses 
at frequencies $\omega_{ab}$ and $\omega_{ac}$. Under the 
condition of adiabatic passage, the molecule in the initial state 
$|b\rangle$ is transformed into a coherent state $(|b\rangle - 
|c\rangle)/\sqrt{2}$. 
 
The principle behind a STIRAP process is the adiabatic theorem 
as applied to the time-varying Hamiltonian $H(t)$. If the 
system at time $t_0$ is in an eigenstate of $H(t_0)$, 
and the evolution from $t_0$ to $t_1$ is sufficiently slow, 
then the system will evolve into the eigenstate of $H(t_1)$. 
The three-level atomic system driven by two fields 
has three eigenstates, one of which is 
a linear superposition of only the lower 
levels $b$ and $c$. The time dependent amplitudes 
of this eigenstate depend on the 
pulse shapes of the fields at frequencies $\omega_{ab}$ and 
$\omega_{ac}$. Thus, by an appropriate pulse shaping, it should 
be possible to prepare a maximally coherent superposition 
of states $b$ and $c$ as shown in Fig. \ref{Fig9}. The expressions for 
the Hamiltonian and the corresponding eigenstates are given in 
Appendix \ref{App-STIRAP}. 
 
 
Comparing different schemes for the preparation of maximal  coherence, we note
that the required laser power is much lower  for the all-resonant scheme, but
in the case of biomolecules, UV  lasers are required.  The far-detuned scheme
will work with more  powerful infrared lasers, up to the point of laser
damage.   As for the comparison of adiabatic and  non-adiabatic regimes, we
should note that the adiabatic scheme  may turn out to be more robust, because
it does not rely on a  particular pulse area and works for inhomogeneous
molecular  ensembles and non-uniform laser beams. 

 
\subsubsection{Femtosecond Pulse Sequences} 
 
\begin{figure}[h] 
\includegraphics[scale=0.5]{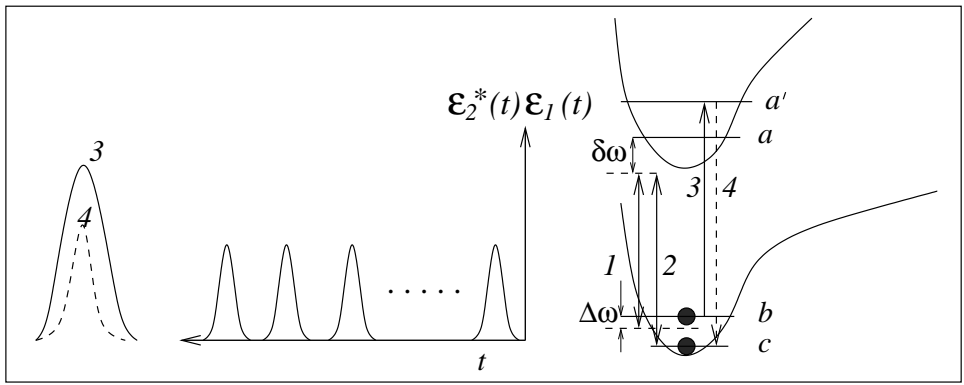} 
\parbox{90ex}{\caption{\label{Fig11a} 
Femtosecond pulse train prepares molecular motion, described by $\varrho_{bc}$.
The molecular state is then probed by field 3 so that field 4 is produced.
\label{Fig10}} }
\end{figure}

In a series of beautiful experiments K. Nelson and coworkers 
\cite{KANelson} have 
generated coherent molecular vibration
via a train of femtosecond pulses, see Fig. \ref{Fig11a}. 
They nicely describe their work as: ``Timed sequences of 
femtosecond pulses have been used to repetitively ``push" 
molecules in an organic crystal\dots, in a manner closely 
analogous to the way a child on a swing may be pushed 
repetitively to reach oscillatory motion." 
 
An interesting aspect of this approach is the fact that the 
individual pulses need not be strong. Only the collective effect 
of many weak pulses is required. This may be helpful if molecular 
``break-up", due to strong ${\cal E}_1$ and ${\cal E}_2$, is a 
problem. This will be further discussed elsewhere. 
 
\subsection{Adaptive Evolutionary Algorithms} 
\label{Subsec-Adaptive} 
 
So far we described how one-photon and two-photon resonant 
pulse sequences can be used to produce a coherent molecular 
superposition state. The idea is that once this state is created, 
a delayed pulse can be applied in order to produce Raman 
scattering which will bear the signature of the molecular system. 
The Raman signal is expected to be optimized when the molecular 
coherence is maximal. 
 
\begin{figure}[h] 
\begin{center} 
\includegraphics[scale=0.5]{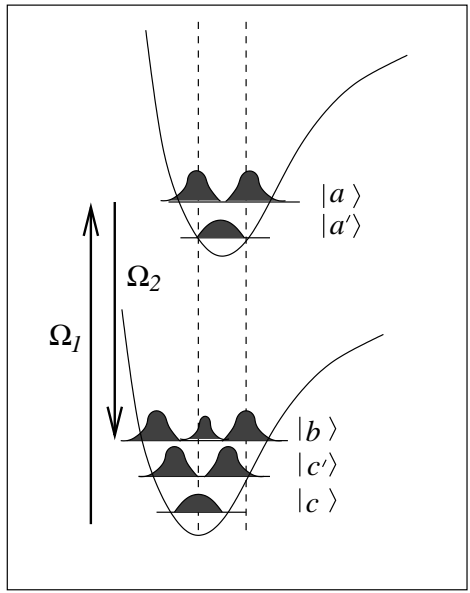}
\end{center} 
\begin{center} 
\parbox{90ex}{\caption{ Schematics 
to generate coherence between the levels $|b\rangle$ 
and $|c\rangle$. The shaded curves represent the probability distribution 
of the interatomic distance for different vibrational and electronic states.
\label{Fig11}} }
\end{center} 
\end{figure}

In general, however, things are complicated by the Franck-Condon 
factors. As an example of a more complicated situation consider 
Fig. \ref{Fig11}. There we see a multilevel system with the ground state 
$|c\ra$ and the next state $|c^{\prime}\ra$. Matrix elements with 
Franck-Condon overlap factors yield a weak transition between 
$|c'\ra$ and $|a\ra$ or $|a'\ra$ as indicated in the figure. 
However, given an appropriately nonlinear ground state potential 
it is quite possible that the next vibrational state could be 
off-set and have the appropriate position of the peaks of the 
wave function in order to maximize the Franck-Condon overlap. In 
this way, the coherence between the states $|b\ra$ and $|c\ra$ of 
Fig. \ref{Fig11} could still serve as a strong generator of the anti-Stokes 
radiation on the $|a\ra$ to $|b\ra$ transitions.

Given such a configuration, it is not hard to see how to prepare 
the ground state coherence. We could, for example, radiate the 
molecule with a chirped Raman pair so as to generate maximal 
coherence between $|b\ra$ and $|c^{\prime}\ra$ and then follow 
that by a Raman $\pi-$pulse transferring the population from 
$|c^{\prime}\ra$ to $|c\ra$. In this way, maximal coherence 
between $|c\ra$ and $|b\ra$ would be prepared.  And as indicated 
in the figure, strong matrix elements would be expected between 
$|b\ra, |c\ra$ and the first excited vibrational state of the 
electronic potential. 
 
The preceding example shows that even in a simplified ideal 
few-level system, preparation of maximal molecular coherence may 
require application of a complicated pulse sequence. For large 
bio-molecules the level structure is not only much more complex, 
but usually unknown. We now consider how search algorithms can be 
used to find the optimal pulse sequence for a complicated molecule 
with an unknown Hamiltonian. This approach will eventually lead 
to an efficient generation of ``molecular fingerprints". 
 
In order to achieve this goal we will need to (1) utilize a 
technique for preparation of complex shaped pulse sequences; (2) 
find the particular pulse sequences, required for the excitation 
of the particular bio-molecules and for the production of 
spectral signatures, which will allow one to distinguish (with 
certainty) the target biological agent from any other species. 
 
Pulse shaping techniques already exist; they are based on 
``spectral modification". First, a large coherent bandwidth is 
produced by an ultra-short pulse generation technique 
\cite{Krausz}. Then, the spectrum is dispersed with a grating or 
a prism, and each frequency component is addressed individually 
by a spatial light modulator (a liquid crystal array 
\cite{Weiner} or an acoustic modulator \cite{Warren}). This way, 
individual spectral amplitudes and phases can be adjusted 
independently. Finally, the spectrum is recombined into a single 
beam by a second dispersive element, and focused onto the target. 
This technique allows synthesis of arbitrarily shaped 
pulses right at the target point, and avoids problems associated 
with dispersion of intermediate optical elements and windows. 
 
A particular shaped pulse sequence can be represented by a 
three-dimensional surface in a space with 
frequency-amplitude-phase axes. Each pulse shape, which 
corresponds to a particular 3-D surface, produces a molecular 
response. The problem is to find the optimal shape. The search 
space is too large to be scanned completely. Besides, many local 
optima may exist in the problem. The solution is offered by 
``global search" algorithms (such as adaptive evolutionary 
algorithms) \cite{Assion,Gerber}. In this approach the 
experimental output is included in the optimization process. This 
way, the molecules subjected to control, are called upon to guide 
the search for an optimal pulse sequence within a learning loop 
\cite{Rabitz}. With the proper algorithm, automated cycling of 
this loop provides a means of finding optimal pulse shapes under 
constraints of the molecular Hamiltonian and the experimental 
conditions. No prior knowledge of the molecular Hamiltonian and 
the potential energy surfaces is needed in this case. 
 
\begin{figure}[h] 
\centerline{\epsfig{file=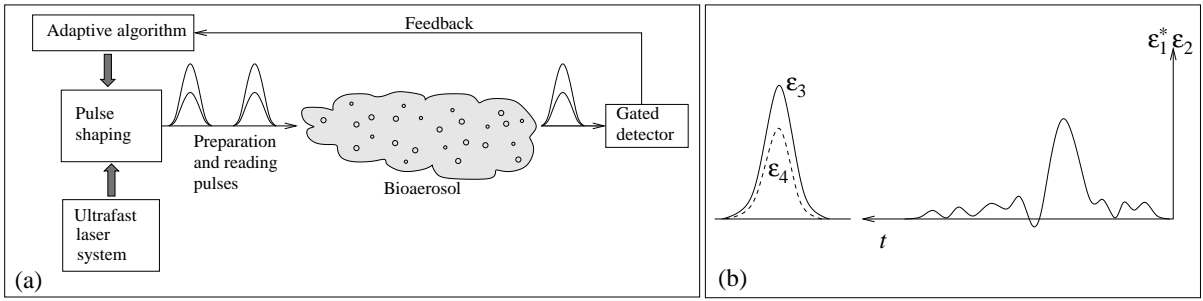,scale=0.3}}%
\begin{center} 
\parbox{90ex}{\caption{\label{Fig13a} 
(a) Experimental setup for the implementation of 
adaptive techniques; 
(b) Figure depicting amplitude of possible 
optimized Raman preparation pulse sequence ${\cal E}^*_1 {\cal E}_2$. 
Not indicated is the fact that the 
``reading'' pulse ${\cal E}_3$ can also be profitably considered as a learning
algorithm variable.
} }
\end{center} 
\end{figure}

This adaptive technique was developed for coherent control of 
chemical reactions \cite{Gerber}. The idea is that the pulses 
can be optimized to produce desired chemical products. In our 
problem we want to optimize Raman generation. In this case both 
preparation and reading pulses can be adaptively shaped in order 
to maximize the signal. Fig. \ref{Fig13a} shows 
schematics for the 
experimental setup that implements these ideas. 
 
Generated spectra will be different for different molecular 
species. And our task is not only to maximize Raman generation, 
but also to identify spectral patterns characteristic of 
particular species and maximize the difference in the spectrum 
produced by the target bio-molecule from spectra produced by any 
other bio-molecules. The key idea here is to apply the same 
adaptive algorithms in order to learn these optimal ``molecular 
fingerprints". 
 
We note that the complexity of the molecular level structure is 
not so much a problem as a solution to a problem. We can take 
advantage of the richness of the molecular structure, and the 
infinite variety of possible pulse shapes, in order to 
distinguish different species with the required certainty. 
\pagebreak


\section{Possible FAST CARS measurement strategies for detection of bacterial 
spores} 
\label{Sec-FAST-CARS} 
 
Having presented the FAST CARS technique in some detail we now return to the 
question of its application to ``fingerprinting'' of macromolecules and 
bacterial spores. Some aspects of the technique seem fairly simple to implement 
and would seem to hold relatively immediate promise. Others are more challenging 
but will probably be useful at least in some cases. Still other applications, 
e.g., the stand-off detection of bioaerosols in the atmosphere present many open 
questions and require careful study. In the following we discuss some simple 
FAST CARS experiments which are underway and/or being assembled in our 
laboratories.

\begin{figure}[h] 
\centerline{\epsfig{file=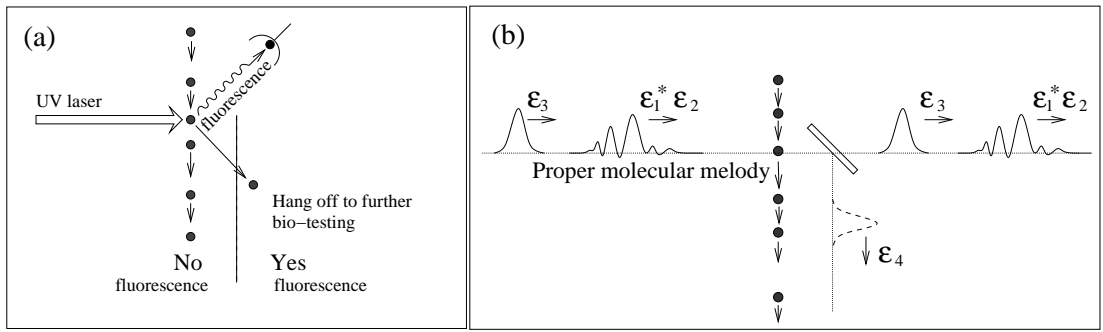,scale=0.35}} 
\begin{center} 
\parbox{90ex}{\caption{(a) 
Two-stage detection scheme. Airborne particles are 
irradiated with a UV laser. Those particles which fluoresce 
are selected for further testing. 
(b) FAST CARS testing of the preselected particles is like ``singing a song'' 
and listening for the standing ovation after an intricate aria. 
That is, the signal is generated only when a particular molecule hears its
molecular melody.
\label{fig-strategy1}} }
\end{center} 
\end{figure} 

 
\subsection{Preselection and hand-off scenarios} 
 
At present, field devices are being engineered which will involve an optical 
preselection stage based on, e.g., fluorescence tagging. If the fluorescence 
measurement does not match  the 
class of particles of interest then that  
particle is ignored. When  many such particles are tested 
and a possible positive is recorded,  the particle is subjected to special 
biological assay; see Fig. \ref{fig-strategy1}a. Such a two stage approach can 
substantially speed up the detection procedure. The relatively simple 
fluorescence stage can very quickly sort out many uninteresting scattering 
centers while the more sophisticated Raman scattering protocol will only be 
used for the captured ``suspects''.

The properly shaped preparation pulse sequence will  be determined by, e.g., 
the adaptive learning algorithm approach as per section \ref{Subsec-Adaptive}. 
The amplitude and phase content of the pulse which produces maximum
oscillation  may be linked to a musical tune. Each spore will have a song which
results in  maximum Raman coherence. A correctly chosen ``melody'' induces a
characteristic  response of the molecular vibrations---a response which is as
unique as  possible for the bacterial spores to be detected.  Playing a melody
rather than  a single tone is a generalization that enables us to see a
multidimensional  picture of the investigated object.  
We note that the optimization can (and frequently will)
include not only the preparation pulses 1 and 2 (see Fig. \ref{Fig13a}b), but
also the probe pulse 3, in particular, its central frequency and timing.
Analysis of the response to such a complex  input is a complicated signal
processing problem. Various data mining  strategies may be
utilized in a way similar to speech analysis.  
 
\begin{figure}[t] 
\centerline{\epsfig{file=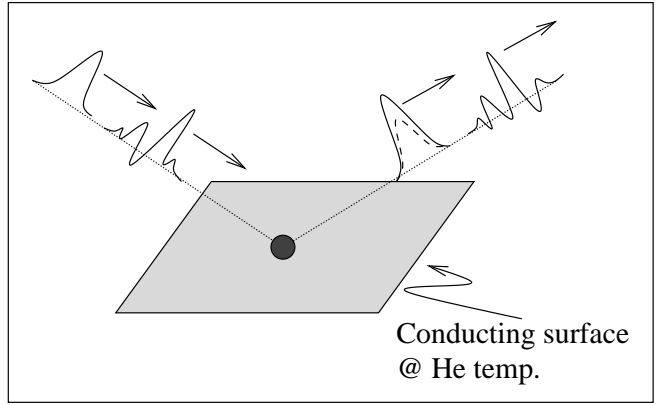,scale=0.5}} 
\begin{center} 
\parbox{90ex}{\caption{Surface Raman payoff: 
Proof of principle we are carrying out, utilizing a cold (He$^4$ 
temperature) sample so as to enhance the Raman signatures and
minimize $\gamma_{bc}$. 
\label{fig-strategy3}} }
\end{center} 
\end{figure} 

However, taking into account the fact
that we work  with femtosecond pulses chained in picosecond to nanosecond pulse
trains, the  whole analysis can be very short. In particular, if we recall the
long sampling  time of the complete fluorescence spectra of \cite{Ghiamati}
being $\approx  15$min, our estimation of a microsecond analysis is a very
strong argument for  the chosen approach.

 
\subsection{Possible further Raman characterization} 
 
After a suspect particle has been targeted, it may be subject to a whole 
variety of investigative strategies. Raman scattering off a flying particle
can  be  very fast, but not necessarily the most accurate method. It will be
very  useful to pin the particle on a fixed surface and cool it down to
maximize the  decoherence time $T_2$ so that the characteristic lines are
narrowed down. The  particle can be deflected by optical means (laser tweezers,
laser ionization,  etc.) and attached to a cooled conducting surface (see Fig. 
\ref{fig-strategy3}). Cooling to liquid helium temperature would enable  us to
enhance the dephasing time from $T_2 \sim 10^{-12}$sec at room temperature  to
$T_2 \lesssim 10^{-9}$sec at a few degrees Kelvin.

\subsection{Possible spore specific FAST CARS  detection schemes} 
 
We conclude with  some speculative observations for long 
range (stand-off) measurements.

\begin{figure}[t] 
\centerline{\epsfig{file=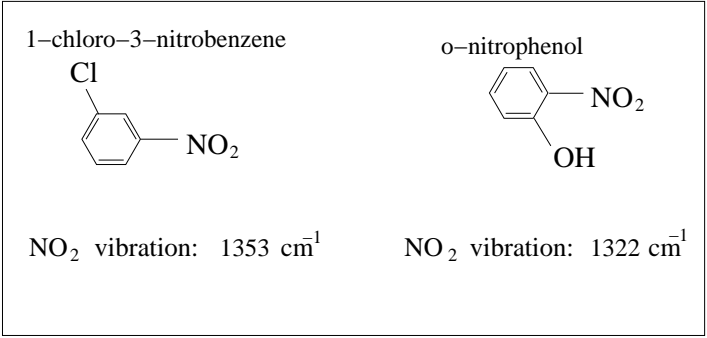,scale=0.35}} 
\parbox{90ex}{\caption{\label{Fig16} Influence of various radicals on the 
frequency of the symmetric NO$_2$ vibrations.} }
\end{figure}

The chemical state of DPA in the spore is of special interest to 
us because the stuff we hang on the DPA molecule will determine 
its characteristic Raman frequency. To this end, we quote from 
article \cite{Murrell} by Murrell on the chemical composition of 
spores: ``When DPA is isolated from spores it is nearly always in 
the Ca-CDPA chelate but sometimes as the chelate of other 
divalent metals [e.g. Zn, Mn, Sr etc.] and perhaps as a DPA-Ca 
amino complex." 
 
Thus,  since each different type of spore would  have its
own unique mixture of metals and amino acids,  it may be the case that the
finer details of the Raman spectra would contain  spore specific
``fingerprints." This conjecture is supported by  Fig. \ref{Fig02}b where the
difference between the DPA Raman spectra of the  spores of {\it Bacillus
cereus} and {\it Bacillus megaterium} is  encouraging. 
 
\begin{figure}[t] 
\centerline{\epsfig{file=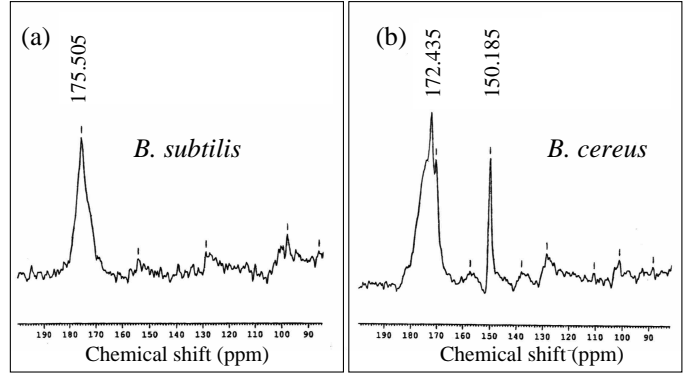,scale=0.45}} 
\parbox{90ex}{\caption{\label{FigNMR} 
Nuclear magnetic resonance $^{13}$C Cross polarization magic angle spinning
(CPMAS) spectrum of (a) outer coatless mutant {\it Bacillus subtilis} 322 spores
and (b) dormant {\it Bacillus cereus} spores. Complete spectra is found in
\cite{Leuschner01}.
} }
\end{figure}

The open question is: to what extent is 
the DPA Raman spectra sensitive to its 
environment? 
That we might be able to achieve spore specific sensitivity is
consistent with the well known fact 
that substituents, e.g., NO$_2$ experience a substantial shift of 
their vibrational frequencies when bound in  different molecular 
configurations,~see~Fig.~\ref{Fig16}. Furthermore, recent NMR experiments 
\cite{Leuschner01} show
spore specific fingerprints due to the local environment
(see Fig. \ref{FigNMR}).

Clearly there are many oportunities and open questions implicit in the FAST
CARS ``molecular melody'' approach to real time spectroscopy. However it plays
out, this combination of quantum coherence and coherent control promises to be a
fascinating area of research.


\acknowledgments

The authors gratefully acknowledge the support from 
Air Force Research Laboratory (Rome, New York), 
DARPA-QuIST, 
TAMU Telecommunication and Informatics Task Force (TITF) Initiative, 
ONR (contract N000014-95-1-0275), 
and the Welch
Foundation. We would also like to thank
R. Allen,
Z. Arp,
A. Campillo,
K. Chapin,
R. Cone,
A. Cotton,
E. Eisenstadt,
J. Eversole,
M. Feld,
J. Golden,
S. Golden,
T. Hall,
S. Harris,
P. Hemmer,
J. Laane,
F. Narducci,
B. Spangler,
W. Warren,
G. Welch,
S. Wolf,
and
R. Zare
for valuable and helpful discussions.

\newpage 
 
\appendix 
\pagebreak 
\renewcommand{\theequation}{A\arabic{equation}} 
  \setcounter{equation}{0}  
 
 
\section{Three level system driven by two fields}  
\label{App-Dream} 
 
In this appendix we consider the density matrix approach 
to the Raman scattering and discuss various limiting cases. 
In particular we present a semiclassical treatment in 
which the field evolution is described by Maxwell's equations 
and the atomic system by the density operator. 
 
We consider a three level atomic system in the $\Lambda$ configuration 
with upper level a and lower levels b and c. The $a\rightarrow b$ 
transition is  driven by a field at frequency $\nu_1$ and the 
$a\rightarrow c$ transition is coupled via a signal field at 
frequency $\nu_2$.

The Maxwell's equations lead to the following equation for the 
evolution of the signal field, 
\begin{eqnarray} 
 \left(\frac{\partial^2}{\partial z^2}-\frac{1}{c^2}\frac{\partial^2}
 {\partial t^2} \right)E 
 &=&\mu_o\ddot{P} , 
 \label{Maxwell1} 
\end{eqnarray} 
where $E$ is the electric field vector of the emitted light, 
$\mu_o$ is the vacuum permeability, and {\cal P} is the medium 
polarization. We write the electric field using the slowly 
varying envelope {\cal E} as 
\begin{eqnarray} 
 E&=&\frac{1}{2}{\cal E}(z,t)e^{-i(\nu t-kz+\phi)}+c.c. 
 \label{E-as-slow} 
\end{eqnarray} 
and write the polarization in terms of the slowly varying 
quantity $P$ as 
\begin{eqnarray} 
 P&=&\frac{1}{2} {\cal P}(z,t)e^{-i(\nu t-kz+\phi)}+c.c. \ . 
 \label{P-as-slow} 
\end{eqnarray} 
Working within the slowly varying amplitude and phase 
approximation, Eqs. (\ref{Maxwell1}), (\ref{E-as-slow}), and 
(\ref{P-as-slow}) yield 
\begin{equation} 
 \frac{1}{c}\frac{\partial {\cal E}}{\partial t}+ 
 \frac{\partial {\cal E}}{\partial z} 
 =-\frac{1}{2\epsilon_0}k\text{Im}{\cal P}, 
\end{equation} where 
\begin{eqnarray} 
 {\cal P}=2 {\cal N}(z,t) \wp_{ab}\varrho_{ab}e^{i(\nu t-kz+\phi)}. 
\end{eqnarray} 
Here $\wp_{ab}$ is the dipole moment matrix element, 
$\varrho_{ab}$ is the off-diagonal element of the density matrix 
for the molecular levels $a$ and $b$, and ${\cal N}(z,t)$ is the 
volume density of the molecules. We note that the electromagnetic 
field is determined by $\varrho_{ab}$.

\begin{figure}[ht] 
\centerline{\epsfig{file=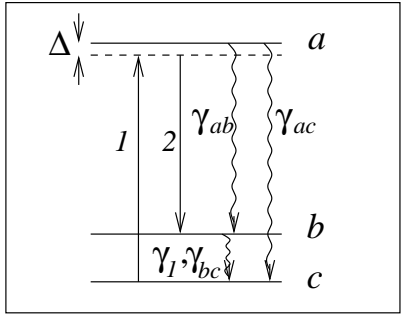,scale=0.5}} 
\parbox{90ex}{\caption{Schematic 
diagram of the Raman interaction for the three-state 
system. The driving fields 1 and 2 are in Raman resonance with the two lower
levels $b$ and $c$, but are generally off-resonant with the transitions to the
excited state $a$ (with detuning $\Delta$). Decoherence and decay of the
optical transitions $ab$ and $ac$ are characterized by the rates $\gamma_{ab}$
and $\gamma_{ac}$, 
the spontaneous decay rate of level $b$ to $c$ is $\gamma_1$, and
the decoherence rate of the $bc$ transition is $\gamma_{bc}$.
\label{FigB-1}} }
\end{figure} 

Here we recall the main results of \cite{Dream} and apply them to 
our physical situation (see Fig. \ref{FigB-1}). 
First we define the decay rates 
\begin{eqnarray} 
 \gamma_{ab} &=& \frac{\gamma +  \gamma_{1}}{2} + \gamma_{ab}^{p}, \\ 
 \gamma_{ca} &=& \frac{\gamma}{2} + \gamma_{ca}^{p}, \\ 
 \gamma_{cb} &=& \frac{\gamma_{1}}{2} + \gamma_{cb}^{p}. 
\end{eqnarray} 
Here the decay rates from $a$ to $b$ is $\gamma_b$, and from $a$ 
to $c$ is $\gamma_c$, and $\gamma = \gamma_b + \gamma_c$. 
Population decay rate    from $b$ to $c$ is $\gamma_1$. 
Purely phase decays are designated by 
the superscript $p$. Complex dephasings are defined as 
\begin{eqnarray} 
 \Gamma_{ab} &=& \gamma_{ab} - i \Delta_{ab} , \\ 
 \Gamma_{ca} &=& \gamma_{ca} + i \Delta_{ac} ,\\ 
 \Gamma_{cb} &=& \gamma_{cb} , 
\end{eqnarray} 
where $\Delta_{ac} = \omega_{ac}-\nu_1$ and 
$\Delta_{ab} = \nu_2 - \omega_{ab}$.
In the following we assume Raman 
resonance, i.e., $\Delta _{ac} = - \Delta_{ab} = \Delta$. 
The main working equations for the 
off-diagonal density matrix elements are 
\begin{eqnarray} 
 \dot \varrho_{ab} &=& - \Gamma_{ab} \varrho_{ab} + i \Omega_2 
 (\varrho_{bb} - \varrho_{aa}) + i \Omega_1 \varrho_{cb}, \\ 
  \dot \varrho_{ca} &=& - \Gamma_{ca} \varrho_{ca} + i \Omega_1^{*} 
 (\varrho_{aa} - \varrho_{cc}) - i \Omega_2^{*} \varrho_{cb}, \\ 
 \dot \varrho_{cb} &=& - \Gamma_{cb} \varrho_{cb} - i \Omega_2 
 \varrho_{ca} + i \Omega_{1}^{*} \varrho_{ab} . 
\end{eqnarray} 
The equations for the populations are 
\begin{eqnarray} 
 \label{dotrhobb}
 \dot \varrho_{bb} &=& i \Omega^{*}_2 \varrho_{ab} - i\Omega_2 \varrho_{ba} 
 + \gamma_{b} \varrho_{aa}   
 -\gamma_{1} \varrho_{bb}, \\ 
 \dot \varrho_{cc} &=& i \Omega^{*}_1 \varrho_{ac} - i\Omega_1 \varrho_{ca} 
 + \gamma_{c} \varrho_{aa}  + \gamma_{1} \varrho_{bb} , 
\end{eqnarray} 
and $\varrho_{aa}$ is obtained from 
\begin{eqnarray} 
 1 &=& \varrho_{aa} + \varrho_{bb} + \varrho_{cc} . 
\end{eqnarray} 
Here $\Omega_{1,2}$ are the Rabi frequencies of the fields 
having frequencies $\nu_{1,2}$, i.e., 
\begin{eqnarray} 
 \label{Omega-Rabi}
 \Omega_1&=&\frac{\wp_{ac}{\cal E}_1}{\hbar}, \\ 
 \Omega_2&=&\frac{\wp_{ab}{\cal E}_2}{\hbar}, 
\end{eqnarray}

The steady state solution of these equations can be obtained by 
setting all time derivatives equal to zero. The result is 
\begin{equation} 
 \fbox{$ \displaystyle 
 \varrho_{ab}=-\frac{i}{\Gamma_{ab}}[\Omega_2(\varrho_{aa}-\varrho_{bb})- 
 \Omega_1\varrho_{cb}], 
 $} 
 \label{raman-rho-ab} 
\end{equation} 
and 
\begin{eqnarray} 
  \varrho_{ca} &=& \frac{i\Omega_1^*}{\cal D} 
  \left[ (\varrho_{aa} - \varrho_{cc}) 
 (\Gamma_{ab}\Gamma_{cb} + |\Omega_1|^2) + (\varrho_{bb} - \varrho_{aa}) 
 |\Omega_{2}|^2 \right] , \\ 
 \varrho_{cb} &=& \frac{\Omega_2 \Omega_1^*}{\cal D} \left[ 
 (\varrho_{aa} - \varrho_{bb}) \Gamma_{ca} 
 + (\varrho_{aa} - \varrho_{cc}) \Gamma_{ab} \right] , 
 \label{ap-coher} 
\end{eqnarray} 
where the common denominator is 
\begin{eqnarray} 
 {\cal D} = \Gamma_{ab} \Gamma_{ca} \Gamma_{cb} + \Gamma_{ab} |\Omega_2|^2 
 + \Gamma_{ca} |\Omega_1|^2 . 
\end{eqnarray} 
 
Equation (\ref{raman-rho-ab}) is our main working equation. The 
first term in the parenthesis of Eq. (\ref{raman-rho-ab}) is 
responsible for the stimulated emission or absorption on the 
$|a\rangle \leftrightarrow |b\rangle$ transition. The second, 
$\varrho_{cb}$, term describes the Raman conversion of $\nu_1$ 
into $\nu_2$. As we can see, it is 
crucial to have the element $\varrho_{cb}$ as big as possible 
for optimum Raman conversion. In 
the following we present the main properties of $\varrho_{cb}$ 
for various experimental and conceptual configurations  and 
consider various limiting cases for the Raman processes. 
 
 
\subsection{Off-resonant Raman process, weak driving}
\label{App-off-weak} 
 
First we consider the Raman effect far from electronic resonance 
($\Delta \gg \gamma_{ab}, \gamma_{ac}$). In this case the upper 
level $a$ is almost completely depopulated and can be eliminated 
from the dynamics. The system behaves basically as a two-level 
system with the Rabi frequency of oscillations between levels $b$ 
and $c$ equal to $\Omega_{R} = |\Omega_{1}^*\Omega_{2}|/\Delta$. 
For dephasing rate $\gamma_{bc} \gg \Omega_{R}$, we find 
the steady state solution for the lower 
state coherence $\varrho_{cb}$ as 
\begin{eqnarray} 
 \lim_{\Delta \gg 
 \gamma} \varrho_{cb} & = & 
 \hskip.25cmi\frac{\Omega^*_1\Omega_2} 
 {\gamma_{bc}\Delta}(\varrho_{cc}-\varrho_{bb}), 
\end{eqnarray} 
which, after substituting into Eq. (\ref{raman-rho-ab}), yields the 
density matrix element responsible for the $\nu_2$ radiation as 
\begin{eqnarray} 
 \varrho_{ab}^{\text{Raman}}&=&-i\frac{\Omega_2}{\gamma_{ab}-i\Delta}\left[ 
 (\varrho_{aa}-\varrho_{bb})-i\frac{|\Omega_1|^2} 
 {\gamma_{bc}\Delta} (\varrho_{cc}-\varrho_{bb})\right]. 
 \label{off-res-raman} 
\end{eqnarray} 
Note that for the amplification of the $\nu_2$ field the 
so-called Raman inversion $\varrho_{cc}-\varrho_{bb}$ is necessary. 
In the simplest case when almost all the molecules are in the lowest 
state $c$, $\varrho_{cc} \approx 1$, $\varrho_{bb} \approx 0$, 
$\varrho_{aa} \approx 0$, Eq. (\ref{off-res-raman}) 
reduces to 
\begin{eqnarray} 
  \varrho_{ab}^{\text{Raman}}&=&-i\frac{\Omega_2}{\Delta}\frac{|\Omega_1|^2} 
 {\Delta \gamma_{bc}} . 
\end{eqnarray}

 
\subsection{Off-resonant Raman process, strong driving}
\label{App-off-strong}  

In the case of strong driving the transition between states $b$ and $c$ can be
saturated so that both these levels are significantly populated
(with virtually no population in the excited state $a$). In particular,
one obtains 
from Eq. (\ref{dotrhobb}) the steady state
populations of the lower levels
\begin{eqnarray}
 \label{rhobb-strong-drive}
 \rho_{bb} = \frac{i\Omega_2^* \rho_{ab} - i \Omega_2 \rho_{ba}}
 {\gamma_1} ,
\end{eqnarray}
and $\rho_{cc} = 1-\rho_{bb}$.
Substituting Eq. (\ref{raman-rho-ab})
into Eq. (\ref{rhobb-strong-drive}), and assuming, 
without loss of 
generality, that the Rabi frequencies for the laser fields are real 
($\Omega_1=\Omega_1^{\ast}, \Omega_2=\Omega_2^{\ast}$)
we find that for the far off-resonant case
($\Gamma_{ab} \approx i \Delta$)
\begin{eqnarray} 
 \rho_{bb} = \frac{2 \Omega_2 \Omega_1 \rho_{bc}} 
 {i \Delta \gamma_1 }. 
 \label{B-17} 
\end{eqnarray} 
Rearranging Eq. (\ref{ap-coher}) and inserting Eq. (\ref{B-17}), under the
condition
\begin{eqnarray} 
 \gamma_{bc} \gg \left| - \frac{|\Omega_1|^2}{i\Delta} 
 + \frac{|\Omega_2|^2}{i\Delta} \right| , 
\end{eqnarray} 
we find that
\begin{eqnarray} 
 \rho_{bc} =  -\frac{\Omega_2 \Omega_1}{i\Delta \Gamma_{bc}} - 
 \frac{4  \Omega_2^2 \Omega_1^2 \rho_{bc}}{\gamma_1 \Delta^2 \Gamma_{bc}} . 
 \label{B-19} 
\end{eqnarray} 
Solving Eq. (\ref{B-19}) for $\rho_{bc}$ we obtain 
\begin{eqnarray} 
 \rho_{bc} = \frac{i  \Omega_2 \Omega_1 }{ \Delta } 
 \left( \gamma_{bc} + \frac{4 \Omega_2^2 \Omega_1^2}{\gamma_1 \Delta^2}\right)
 ^{-1} . 
\end{eqnarray} 
It can be shown that the maximum magnitude of the coherence $|\rho_{bc}|$ is 
given by 
\begin{eqnarray} 
 \left| \rho_{bc} \right| = \sqrt{\frac{\gamma_1}{16 \gamma_{bc}}} . 
\end{eqnarray} 
This maximum occurs when 
\begin{eqnarray} 
 \frac{\Omega_1 \Omega_2}{\Delta} = 
 \frac{\sqrt{\gamma_1 \gamma_{bc}}}{2}. 
\end{eqnarray}

 
\subsection{Resonant Raman process, weak driving} 
\label{App-res-weak} 
 
In the resonant case ($\Delta = 0$), the Raman effect has much in 
common with the scheme of lasing without inversion (LWI; see, 
e.g., \cite{Dream}). The lower state coherence is then found to be 
\begin{equation} 
 \varrho_{cb}\stackrel{\Delta=0}{\longrightarrow} 
 \hskip0.25cm\Omega_2\Omega^{*}_1 
 \frac{[(\varrho_{aa}-\varrho_{bb})\gamma_{ca}+(\varrho_{aa}-\varrho_{cc}) 
 \gamma_{ab}]}{\gamma_{ab}\gamma_{ac} 
 \gamma_{cb}+\gamma_{ca}|\Omega_1|^2+\gamma_{ab}|\Omega_2|^2} . 
\end{equation} 
Polarization responsible for the $\nu_2$ radiation is then 
governed by 
\begin{equation} 
 \varrho_{ab}=-i\frac{\Omega_2}{\gamma_{ab}} 
 \left[ (\varrho_{aa}-\varrho_{bb}) 
\left( 1-\frac{|\Omega_1|^2 \gamma_{ca}}{D} \right) 
- (\varrho_{aa}-\varrho_{cc})\frac{|\Omega_1|^2\gamma_{ab}}{D} \right], 
\end{equation} 
where $D=\gamma_{ab}\gamma_{ca}\gamma_{cb}+ 
\gamma_{ab}|\Omega_2|^2+\gamma_{ca}|\Omega_1|^2$.

In the weak driving limit 
($\Omega_{1,2}\ll \gamma_{ab},\gamma_{ac},\gamma_{bc}$), almost all 
the population remains in state $c$ ($\varrho_{cc} \approx 1$, 
$\varrho_{bb} \approx 0$, 
$\varrho_{aa} \approx 0$). The atomic coherence of the lower levels is 
\begin{equation} 
\rho_{cb}=-\frac{\Omega_1^*\Omega_2}{\gamma_{ac}\gamma_{bc}}, 
\end{equation} 
and the corresponding 
expression for the density 
matrix element  $\rho_{ab}$ is 
\begin{equation} 
  \varrho_{ab}^{\text{Resonant Raman}} = 
   -i \frac{\Omega_2}{\gamma_{ab}} 
\frac{|\Omega_1|^2}{\gamma_{ac} \gamma_{bc}} . 
\end{equation}

\subsection{Resonant Raman process, strong driving} 
\label{App-res-strong}
 
In the resonant case we can find the steady state solution of a 
strongly driven three-level system ($|\Omega_{1,2}| \gg \gamma_1, 
\gamma_{bc}$)
in the form of the so-called 
dark state. It is useful to consider two superpositions of the 
lower states $|b\rangle$ and $|c\rangle$ defined as 
\begin{eqnarray} 
 |B\rangle&=&\frac{\Omega_2|b\rangle+\Omega_1|c\rangle}{\Omega},\\ 
 |D\rangle&=&\frac{\Omega_1|b\rangle-\Omega_2|c\rangle}{\Omega} , 
 \label{dark-D} 
\end{eqnarray} 
where $\Omega = \sqrt{|\Omega_1|^2 + |\Omega_2|^2}$. The state 
$|D\rangle$ is completely decoupled from the upper state 
$|a\rangle$ ($\hat H |D\rangle = 0$), and is called the dark 
state. If the molecule starts in any state different from 
$|D\rangle$, the fields $\Omega_1$ and $\Omega_2$ will promote it 
to the upper state $|a\rangle$, which decays to the lower states 
by spontaneous emission. If the molecule is in $|D\rangle$ it 
stays there unchanged. In this way the dark component of the 
state increases and finally the system will end up completely in 
the state $|D\rangle$. 
The populations of the levels $a$ and $b$ are $\rho_{aa}=0$ and 
$\rho_{bb} = |\Omega_1|^2/\Omega^2$, and
the dark state coherence of the lower 
states is, as given by Eq. (\ref{dark-D}), 
\begin{eqnarray} 
 \varrho_{cb}&=&-\frac{\Omega^{*}_1\Omega_2}{\Omega^2}. 
\end{eqnarray} 
Maximum Raman coherence is achieved with $|\Omega_1| = |\Omega_2|$ so that
$|\rho_{bc}^{\text{max}}| = 1/2$.
The element $\varrho_{ab}$ [see Eq. (\ref{raman-rho-ab})] responsible
for the radiation
is given by 
\begin{eqnarray} 
\varrho^{\text{Dark Raman}}_{ab}&=&-\frac{i} 
{\gamma_{ab}}\left[\Omega_2(\varrho_{aa}-\varrho_{bb})+
\frac{|\Omega_1|^2\Omega_2}{\Omega^2}\right] = 0, 
\end{eqnarray} 
i.e., there is no radiation from the dark state. To obtain Raman signal one can,
e.g, switch off the field
$\Omega_2$ after reaching the maximum coherence
$|\varrho_{cb}^{\text{max}}| = 1/2$.  
Then the density matrix element 
$\varrho_{ab}$ goes as 
\begin{eqnarray} 
  \varrho_{ab}^{\text{Max coherence}} = i \frac{\Omega_1}{2\gamma_{ab}}  . 
\end{eqnarray} 
 
To summarize the main results of this appendix, we display the 
density matrix elements responsible for the signal field 
generation in  Table \ref{Tab1}.\\

\pagebreak 
\renewcommand{\theequation}{B\arabic{equation}} 
\setcounter{equation}{0}  

\section{Cooperative Spontaneous Emission} 
\label{App-Dicke}

Since the DPA is contained in the small   ($V\sim (1\mu)^3$) volume 
of the core and since the number of participating DPA molecules 
is huge ($N \gtrsim10^9$) the possibility of Dicke superradiance 
should be carefully considered. At the outset, we should note 
that the short dephasing times of vibronic levels ($T_2$ is in 
the picosecond range) and the unknown inhomogeneous dephasing time 
$T_2^\ast$ will tend to wash out the effect. Still, if the 
spontaneous emission time ($\tau\sim1-10$ nanoseconds) is reduced 
by even a small fraction of N, then cooperative spontaneous 
emission may be important. 
 
Motivated by the proceeding we now turn to a short review of the 
effect from the present, FAST CARS, perspective. We focus on two 
levels $|a\ra$ and $|b\ra$. The previous pulses $\varepsilon_1$ 
and $\varepsilon_2$ beat the coherence between $|b\ra$ and 
$|c\ra$ so that at time $t=0$ 
 
\begin{equation} 
\label{C-1} 
|\Psi(0)\ra=B|b\ra +C|c\ra, 
\end{equation} 
the third pulse, arriving at time T, promotes $|c\ra\rightarrow 
|a\ra$ so that at time $t=\tau$ 
\begin{equation} 
\label{C-2} 
|\Psi(\tau)\ra=A|a\ra+B|b\ra +C^{\prime}|c\ra. 
\end{equation} 
Hence the atoms at time $\tau$ are in a coherent superposition 
$|a\ra$ and $|b\ra$ and this affects the rate of spontaneous 
emission from the N molecules involved. 
 
For the present purposes it is best to start with the Hamiltonian 
describing the interaction of N two level systems ($|a\ra$ and 
$|b\ra$) and the quantized radiation field given by 
\begin{equation} 
\label{C-3} 
H_{\text{int}}=\Sigma_{i,k} \hbar g a^+_k\sigma_ie^{ik\cdot r_i}+adj 
\end{equation} 
where, as discussed, e.g., in \cite{ScullyZubairy}, 
$g_k$ is the coupling 
frequency between the molecules and the $k^{th}$ mode of the 
field, $a^+_k$ is the usual creation operator and 
$\sigma_i=(|b\ra \langle a|)_i$ is the lowering operator for the 
$i^{th}$ molecule. 
 
Consider next the case in which all molecules are located at $r_i 
= r_0$ in a volume small compared to the radiation wavelength 
$\lambda$, and noting that the coupling constant is a slowly 
varying function of $k$ we may write 
\begin{equation} 
\label{C-4} 
H_{\text{int}}=\left[g_ke^{ik_0\cdot 
r_0}\sum_ka^+_k \right]\sum_i\sigma_i, 
\end{equation} 
The important point being that the lowering operator described by 
Eq. (\ref{C-4}) is symmetric in the molecular lowering operator 
$\sigma_i$. 
 
To glean the physics from Eq.  (\ref{C-4}) it is enough to consider 3 
molecules which are all started in their upper state, so that 
\begin{equation} 
\label{C-5} 
|\Psi_{3}\ra=|a_1 a_2 a_3\ra. 
\end{equation} 
 
The state (\ref{C-5}) evolves under the influence of the field according 
to the interaction (\ref{C-4}), and the molecular state develops into 
$|\Psi_{3}\ra  \rightarrow \alpha|\Psi_3\ra + \beta|\Psi_2\ra$ 
where $\alpha$ and $\beta$ are uninteresting constants and 
\begin{equation} 
\label{C-6} 
|\Psi_2\ra=\sum_i\sigma_i|\Psi_3\ra=|b_1a_2a_3\ra + |a_1b_2a_3\ra 
+|a_1a_2b_3\ra. 
\end{equation} 
 
\begin{figure}[h] 
\centerline{\epsfig{file=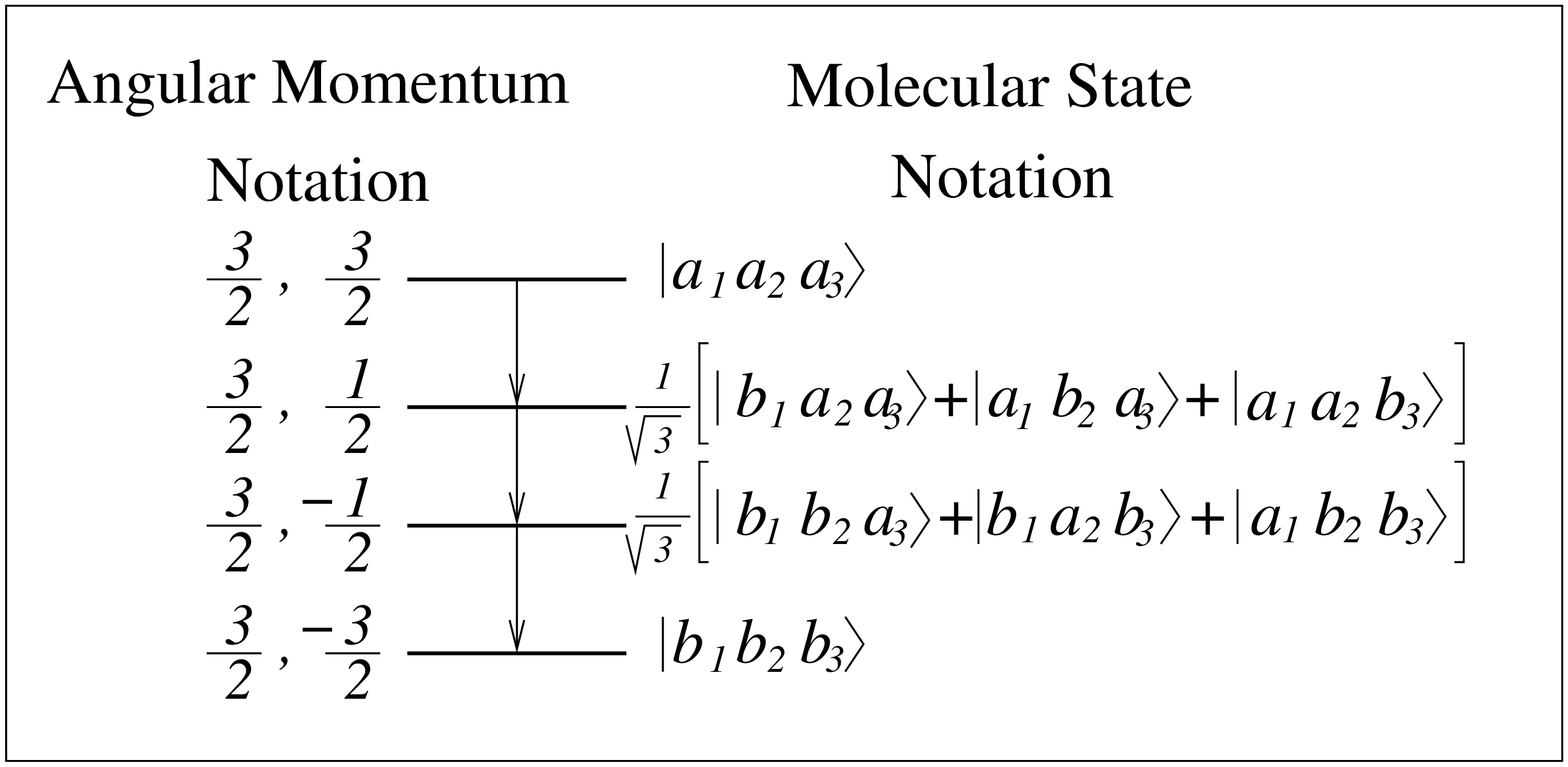,scale=0.35}} 
\parbox{90ex}{\caption{ 
\label{fig-c1} 
Showing that the symmetric combinations of the molecular states $|a_i\ra$ and $|b_i\ra$, i=1,2,3 are 
just the angular momentum states for $j=\frac{3}{2}, 
m_j=\pm\frac{3}{2}, \pm \frac{1}{2}$. } }
\end{figure}

In general the symmetric interaction (\ref{C-4}) only couples the 
$|a_1a_2a_3\ra$ state to the symmetric states of Fig. \ref{fig-c1}. 
 
Thus we may, following Dicke describe our N ``spins" by an 
effective angular momentum of magnitude $j=\frac{1}{2}(N_++N_-)$ 
and projection $m_j=\frac{1}{2}(N_+-N_-)$. 
 
Using this ``angular momentum" picture and recalling that the 
matrix element governing the transition rate between state 
$|j,m_j\ra$ and $|j,m_j+1\ra$ is given by 
\begin{equation} 
\label{C-7} 
\mu=\langle j, m_j+1|\hat{J_-}|j,m_j\ra=\sqrt{(j-m_j)(j-m_j+1)} 
\end{equation} 
we have two interesting limits. 
 
First the case when all spins are up i.e., $N_+=N, N_-=0$ (which 
corresponds to the case of all molecules in $|a\ra$); then 
$j=\frac{N_++N}{2} \rightarrow \frac{1}{2}$ and $m_j = 
\frac{N_+-N}{2} \rightarrow \frac{N}{2}$ and we have 
 
\begin{equation} 
\label{C-8} 
\mu=\langle j,m_j+1|\hat{J_-}|j,m_j\ra = \sqrt{N}. 
\end{equation} 
Therefore the radiation rate goes as $\mu^2=N$, which is the 
normal result for N independent radiators. 
 
In the other case suppose that $m_j=0$ so that $N_+=\frac{N}{2}$ 
and $N_-=\frac{N}{2}$. Then the relevant matrix element is 
\begin{equation} 
\label{C-9} 
\mu=\langle j,-1|\hat{J_-}|j,0\ra =\sqrt{j(j+1)} \cong N, 
\end{equation} 
and the radiation rate goes as $N^2$. That is the state $|j,0\ra$ 
decays N times faster than the state $|j,j\ra$; and therefore the 
state for which $m_j=0$ (equivalently $N_+=\frac{N}{2}$ and 
$N_-=\frac{N}{2}$) is said to be ``superradiant." In fact, any 
such state for which $|m_j| \ll |j|$ is said to be superradiant. 
 
Returning now to our problem in which the levels $|a\ra$ and 
$|b\ra$ are coherently prepared, suppose that A and B of Eq. (\ref{C-1}) 
are both $\frac{1}{\sqrt2}$. An ensemble of N such coherently 
prepared molecules will be described by a superposition of $m_j$ 
levels sharply peaked about $m_j=0$ and is thus superradiant. 
 
The major potential payoff from our perspective is the hope that 
the coherent resonant Raman process may direct the molecules from 
$c\rightarrow a \rightarrow b$ on a time scale which is of the 
order $\tau/N$ where $\tau$ is the spontaneous fluorescence time. 
This would be interesting since, as emphasized by Nelson and 
coworkers \cite{Manoharan90}: ``This [utilization of resonant Raman 
spectra] is possible, however, only in the absence of 
fluorescence interference." 
 
But, as is discussed in Section  \ref{Sec-FAST-CARS} 
there are many open questions. 
For example, will the coherence decay rates $T_2$ and $T_2^\ast$ 
spoil the effect? These and other questions need to be addressed 
by careful experiments and further analysis. 
 
 
\pagebreak 
\renewcommand{\theequation}{C\arabic{equation}} 
\setcounter{equation}{0}  
 
\section{Eigenstates and eigenvectors for a molecular system 
with large one-photon detunings.} 
 
\label{App-Eigenstates} 
 
In the case of large one-photon detunings $\delta \omega$ 
(such that only small virtual excitation of the upper state 
is possible), the molecular system of Fig. \ref{figprepcoher2} 
is described by an effective two-by-two Hamiltonian 
\cite{CombGeneration}: 
\begin{eqnarray} 
\label{Hamilton} 
H_{\rm eff} 
 = - \frac{\hbar}{2} 
\left[ 
\begin{array}{cc} 
A \ & \Omega \\ 
\Omega^* \ & D - 2 \Delta \omega
\end{array} 
\right] ,
\end{eqnarray} 
where $\Omega = \Omega_b \Omega_c/ \delta \omega$ 
is the effective Raman Rabi frequency, 
$A$ and $D$ are dynamic Stark shifts
(see \cite{CombGeneration} for more details),
and $\Delta \Omega$ is the (two-photon) Raman detuning. 
The eigenvalues of the Hamiltonian (\ref{Hamilton}) are
\begin{eqnarray}
 E_{\pm} = \frac{-(A+D-2\Delta \omega) \pm 
 \sqrt{(A-D+2\Delta \omega)^2} + 4 |\Omega |^2}{4},
\end{eqnarray}
and the corresponding eigenvectors are
\begin{eqnarray}
 \label{eqC2}
 |\pm \rangle = \frac{- \Omega |b\rangle + (A + 2E_{\pm}) |c \rangle}
 {\sqrt{|\Omega|^2+(A+2E_{\pm})^2}} .
\end{eqnarray}
In the case when $A \approx D$ we have
\begin{eqnarray}
 E_{\pm} \approx - \frac{A-\Delta \omega \mp \sqrt{\Delta \omega^2 + |\Omega|^2}}
 {2}.
\end{eqnarray}
The probabilities of occupation of states $|b\rangle$ and $|c\rangle$
are
\begin{eqnarray}
 P_b &=& \frac{|\Omega|^2}{|\Omega|^2 + (A+2E_{\pm})^2} \nonumber \\
 ~ & \approx & \frac{|\Omega|^2}{2 \left( |\Omega|^2 + \Delta \omega^2
 \pm \Delta \omega \sqrt{\Delta \omega^2 + |\Omega|^2} \right)} , \\
 P_c & \approx &  \frac{\left( \Delta \omega \pm
 \sqrt{\Delta \omega^2 + |\Omega|^2}\right)^2}
 {2 \left( |\Omega|^2 + \Delta \omega^2
 \pm \Delta \omega \sqrt{\Delta \omega^2 + |\Omega|^2} \right)} .
\end{eqnarray} 
Thus, if we start at $t\to - \infty$  on eigenstate $|-\rangle$ with $\Delta
\omega \to - |\Delta \omega|$, we have $P_b(-\infty)=0$ and  $P_c(-\infty) =
1$. Chirping $\Delta \omega$ such that for $t\to +\infty$ we have $\Delta
\omega \to  |\Delta \omega|$, the $|-\rangle$ eigenstate changes adiabatically
such  that the
probabilities get exchanged to  $P_b(\infty)=1$ and  $P_c(\infty) = 0$.
At the moment when $\Delta \omega = 0$ we have $P_b = P_c = 1/2$ and
$\rho_{bc} = -1/2 e^{i\phi}$, where $\Omega = |\Omega| e^{i\phi}$. Thus, if we
turn off the pulses at this time, we reach full coherence.

 
\pagebreak 
\renewcommand{\theequation}{D\arabic{equation}} 
\setcounter{equation}{0}  
 
\section{Fractional STIRAP} 
 
\label{App-STIRAP} 
 
We consider a three-level atomic system driven by resonant pulses 
at frequencies $\omega_{ab}$ and 
$\omega_{ac}$ with Rabi frequencies 
$\Omega_1(t)=\wp_{ab} {\cal E}_1(t)/\hbar$ and 
$\Omega_2(t)=\wp_{ac}{\cal E}_2(t)/\hbar$, respectively
(se \cite{vss} for more details).  The 
Hamiltonian for the system in the 
slowly varying amplitude and phase approximation is 
\begin{equation} 
H = \frac{\hbar}{2} ( \Omega_1 (t)|a \rangle \langle b| + \Omega_2 
(t)|a\rangle \langle c| + H.c.), 
\end{equation} 
where $H.c.$ means Hermitian conjugate and $|a\rangle$ is the 
excited state. This system has a dark state. The eigenvalue of 
the Hamiltonian for this state is equal to zero, $\lambda_{D}=0$, 
i.e., $\hat H |D\rangle = \lambda_{D} |D\rangle = 0$.  For the 
scheme shown, the dark state is 
\begin{equation} \label{d1} 
|D\rangle = \frac{\Omega_1 (t)|c\rangle - \Omega_2 (t) |b\rangle} 
{ \sqrt{|\Omega_1(t)|^2 + |\Omega_2(t)|^2 }}. 
\end{equation} 
This state mixes the ground states $|b\rangle$ and $|c\rangle$ and 
is independent of the excited state $|a\rangle$.

The sequence of pulses with Rabi frequencies $\Omega_1(t)$ and  $\Omega_2(t)$
is such that, for the molecular system initially in  the state $|b\rangle$,
$\Omega_1(t)/\Omega_2(t)\rightarrow 0$ at  $t\rightarrow -\infty$ and
$\Omega_1(t)/\Omega_2(t)\rightarrow  {\rm tan}\theta$ as $t\rightarrow
+\infty$. Thus the   state evolves adiabatically from $|b\rangle$ to the
coherent  state ${\rm cos}  \theta|b\rangle - {\rm sin}\theta |c\rangle$. For
the case when  $\Omega_1(t)/\Omega_2(t)\rightarrow 1$, $\theta= \pi/ 4$ and
the  resulting state is  $(|b\rangle - |c\rangle)/\sqrt{2}$. We can see that,
unlike the case of conventional STIRAP, the two pump pulses vanish {\em
simultaneously}. In the conventional STIRAP the main aim is to bring a system
from one low-lying energy eigenstate to another without populating an excited
state on the way. The pump pulses in the conventional STIRAP are timed such
that the pulse connected to the target state starts before the beginning and
ends before the end of the pulse connected to the initial state (i.e., the so
called counterintuitive sequence).  The main aim of the fractional STIRAP is to
bring the system from an energy eigenstate to a preselected superposition of
two lower lying eigenstates, without populating any other state. This task
requires a modified timing of the pump pulses. In contrast to the scheme
discussed in the preceding Appendix \ref{App-Eigenstates}, the pump pulses of
the fractional STIRAP are resonant which means that much weaker field or
shorter duration of the pulses is required.


\newpage
 
 
\begin{thebibliography}{99} 
 
 
 
\bibitem{Seaver99} 
M. Seaver, J.D. Eversole, J.J. Hardgrove, W.K. Cary, and D.C. Roselle, 
Aerosol Sci. Tech. {\bf 30,} 174 (1999). 
 
\bibitem{Cheng} 
Y.S. Cheng, E.B. Barr, B.J. Fan, P.J. Hargis, Jr., D.J. Rader, 
T.J. O'Hern, J.R. Torczynski, G.C. Tisone, B.L. Preppernau, S.A. 
Young, and R.J. Radloff, Aerosol Sci. Tech. {\bf 30,} 186 (1999). 
 
\bibitem{Manoharan90} 
R. Manoharan, E. Ghiamati, R.A. Dalterio, K.A. Britton, W.H. Nelson, and J.F. 
Sperry, 
J. Microbiol. Methods {\bf 11,} 1 (1990). 
 
\bibitem{Nelson91}
W.H. Nelson and J.F. Sperry,
\textit{Modern Techniques in Rapid Microorganism Analysis,} 
edited by W.H. Nelson (VCH Publishers, N.Y. 1991).
 
 
\bibitem{Ghiamati} 
E. Ghiamati, R. Manoharan, W.H. Nelson, and J.F. Sperry, 
Applied Spectroscopy {\bf 46,} 357 (1992). 
 
\bibitem{Manoharan93} 
R. Manoharan, E. Ghiamati, S. Chadha, W.H. Nelson, and J.F. 
Sperry, 
Appl. Spectroscopy {\bf 47,} 2145 (1993). 

\bibitem{Terror} 
\textit{Chemical and Biological Terrorism: Research and Development to 
Improve Civilian Medical Response.}
(National Academy Press, Washington, D.C., 1999), p. 90. 
 
\bibitem{ScullyZubairy}
M.O. Scully, Phys. Rev. Lett. {\bf 67,} 1855 (1991);
M.O. Scully, Phys. Rep. {\bf 219,} 191 (1992);
M.O. Scully and M.S. Zubairy,
\textit{Quantum Optics} (Cambridge University Press, Cambridge, 1997). 
 
\bibitem{LWI} 
O. Kocharovskaya and Ya. I. Khanin,
Pis'ma Zh. Eksp. Teor. Fiz. {\bf 48,} 581 (1988) (JETP Lett. {\bf 48,} 630
(1988));
S.E. Harris, Phys. Rev. Lett. {\bf 62,} 1033 (1989);
M.O. Scully, S.Y Zhu, and A. Gavrielides, Phys. Rev. Lett {\bf 
62,}  2813 (1989); A.S. Zibrov, M.D. Lukin, D.E. Nikonov, L. 
Hollberg,  M.O. Scully, V.L. Velichansky, and H.G. Robinson, ibid 
{\bf 75,}  1499 (1995). 
 
\bibitem{EIT} 
K.--J. Boller, A. Imamo\u{g}lu, and S.E. Harris, Phys. Rev. Lett. {\bf 66,} 2593 
(1991); J.E. Field, K.H. Hahn, and S.E. Harris, Phys. Rev. Lett. {\bf 67,} 3062 
(1991). 
 
\bibitem{slowlight} 
L. V. Hau, S.E. Harris, Z. Dutton, and C.H. Behroozi, 
Nature {\bf 397,} 594 (1999); M.M. Kash, V.A. Sautenkov, A.S. 
Zibrov, L. Hollberg, G.R. Welch, M.D. Lukin, Y. Rostovtsev, E.S. 
Fry, and M.O. Scully, Phys. Rev. Lett. {\bf 82,} 5229 (1999); O. 
Kocharovskaya, Yu. Rostovtsev, and M.O. Scully, Phys. Rev. Lett. 
{\bf 86,} 628 (2001); 
M. Fleischhauer and M.D. Lukin, Phys. Rev. Lett. {\bf 84,} 5094 (2000); 
C. Liu, Z. Dutton, C.H. Behroozi, and L. V. Hau, 
Nature {\bf 409,} 490 (2001); D.F. Phillips, A. Fleischhauer, A. Mair, R.L. 
Walsworth, and M.D. Lukin,  Phys. Rev. Lett. {\bf 86,} 783 (2001). 
 
\bibitem{sokolov} 
J.Q. Liang, M. Katsuragawa, F.L. Kien, and K. Hakuta,
Phys. Rev. Lett. {\bf 85,} 2474 (2000).
A.V. Sokolov, D.R. Walker, D.D. Yavuz, G.Y. Yin, and S.E. Harris, 
Phys. Rev. Lett.  {\bf 87,} 3402 (2001). 
 
\bibitem{Rabitz} R. S. Judson and H. Rabitz, Phys. Rev. Lett. {\bf 68}, 
1500 (1992). 
  
 
\bibitem{Kosloff}
R. Kosloff, S.A. Rice, P. Gaspard, S. Tersigni, and D.J. Tannor,
Chem. Phys. {\bf 139,} 201 (1989).

\bibitem{Warren93}
W.S. Warren, H. Rabitz, and M. Dahleh,
Science {\bf 259,} 1581 (1993).

\bibitem{Gordon97}
R.J. Gordon and S.A. Rice,
Annu. Rev.  Phys. Chem. {\bf 48,} 601 (1997).

\bibitem{Zare}
R.N. Zare,
Science {\bf 279,} 1875 (1998). 
 
\bibitem{Rabitz00} 
H. Rabitz, R. de Vivie-Riedle, M. Motzkus, and K. Kompa,
Science {\bf 288,} 824 (2000).

\bibitem{Brixner01}
T. Brixner, N.H. Damrauer, and G. Gerber,
Advances At. Molec. Opt. Phys. {\bf 46,} 1 (2001).

\bibitem{Brumer86}
P. Brumer and M. Shapiro,
Chem. Phys. Lett. {126,} 541 (1986).

\bibitem{Tannor86}
D.J. Tannor, R. Kosloff, and S.A. Rice,
J. Chem. Phys. {\bf 85,} 5805 (1986).

\bibitem{Bergmann98}
K. Bergmann, H. Theuer, and B.W. Shore,
Rev. Mod. Phys. {\bf 70,} 1003 (1998). 

\bibitem{Heritage}
J.P. Heritage, A.M. Weiner, and R.N. Thurston,
Opt. Lett. {\bf 10,} 609 (1985).

\bibitem{Weiner88}
A.M. Weiner, J.P. Heritage, and E.M. Kirschner,
J. Opt. Soc. Am. B {\bf 5,}  1563 (1988).


\bibitem{Wefers95}
M.M. Wefers and K.A. Nelson,
Opt. Lett. {\bf 20,} 1047 (1995).

\bibitem{Weiner00}
A.M. Weiner,
Rev. Sci. Instrum. {\bf 71,} 1929 (2000).

\bibitem{Demtroder}
W. Demtr\"{o}der,
\textit{Laser Spectroscopy,}
(Springer, Berlin 1981).

\bibitem{Heid01}
M Heid, S. Schlucker, U. Schmitt, T. Chen, R. Schweitzer-Stenner, 
V. Engel, and W. Kiefer,
J. Raman Spect. {\bf 32,} 771 (2001).

\bibitem{Chen00}
T. Chen, A. Vierheilig, P. Waltner, M. Heid, W. Kiefer, and A.
Materny,
Chem. Phys. Lett. {\bf 326,} 375 (2000).

\bibitem{Zeidler02}
D. Zeidler, S. Frey, W. Wohlleben, M. Motzkus, F. Busch,
T. Chen, W. Kiefer, and A. Materny,
J. Chem. Phys. {\bf 116,} 5231 (2002).

\bibitem{JRS00}
J. Raman Spectr. {\bf 31,} number 1/2, ed. W. Kiefer (2000).

\bibitem{Silberberg}
D. Oron, N. Dudovich, D. Yelin, and Y. Silberberg,
Phys. Rev. Lett. {\bf 88,} 063004 (2002); N. Dudovich, D. Oron, and Y.
Silberberg, ibid. {\bf 88,} 123004 (2002).



\bibitem{Black} 
J. Black {\it Microbiology: Principles and Explorations} (J. 
Wiley Publishers, 2002). 

\bibitem{Talaro}
K. Talaro and A. Talaro, {\it Foundations in Microbiology}, p. 80
(William C. Brown Publishers, 1999).


 
\bibitem{Smekal} 
A. Smekal, Naturwissenschaften {\bf 11,} 873 (1923). 
 
\bibitem{Compton} 
A. Compton, Phys. Rev. {\bf 21,} 483 (1923). 
 
\bibitem{Raman} 
C.V. Raman and K.S. Krishnan, Nature {\bf 121,} 501 (1928). 
 
\bibitem{Landsberg} 
G. Landsberg and L. Mandelstam, Naturwissenschaften {\bf 16,} 557 
(1928). 
 
\bibitem{Boyd} 
R.W. Boyd, \textit{Nonlinear Optics} (Academic Press, Boston 
1992). 
 
\bibitem{Dream} 
D.E. Nikonov, M.O. Scully, M.D. Lukin, E.S. Fry, L.W. Hollberg, 
G.G. Padmbandu, G.R. Welch, and A.S. Zibrov, Proceedings of 
``Coherent phenomena and amplification without inversion, St. 
Petersburg, 1995", SPIE vol. 2798, 342 (1996). 
 
 
\bibitem{CombGeneration} S. E. Harris and A. V. Sokolov, 
Phys. Rev. A {\bf 55}, R4019 (1997). 
 
\bibitem{Grischkowsky}  D. Grischkowsky, Phys. Rev. Lett. 
{\bf 24}, 866 (1970); D. Grischkowsky and 
\hbox{J.~A.~Armstrong,}  Phys. Rev. A {\bf 6}, 1566 (1972). 
 
\bibitem{Oreg}  J. Oreg, F. T. Hioe, and J. H. Eberly, Phys. Rev. A 
{\bf 29}, 690 (1984). 
 
\bibitem{Kaplan}  A. E. Kaplan, Phys. Rev. Lett. {\bf 73}, 1243 (1994); 
A. E. Kaplan and P. L. Shkolnikov, J.~Opt. Soc. Am. B {\bf 13}, 
347 (1996). 
 
 
\bibitem{vss} N. V. Vitanov, K.-A. Suominen, and B. W. Shore, 
J. Phys. B: At. Mol. Opt. Phys. {\bf 32}, 4535 (1999). 
 
\bibitem{Jain} 
M. Jain, H. Xia, G. Y. Yin, A. J. Merriam, and S. E. Harris, 
Phys. Rev. Lett. {\bf 77}, 4326 (1996). 
 
\bibitem{KANelson}
A.M. Weiner, D.E. Leaird, G.P. Wiederrecht, K.A. Nelson,
Science {\bf 247,} 1317 (1990). 
 
\bibitem{Krausz} The shortest optical pulses generated to date (5-6~fs) 
are obtained by expanding the spectrum of a mode-locked laser by 
self-phase modulation in an optical waveguide, and then 
compensating for group velocity dispersion by diffraction grating 
and prism pairs: R. L. Fork, C. H. Brito-Cruz, P. C. Becker, and 
C. V. Shank, Opt. Lett. {\bf 12}, 483 (1987); A. Baltuska, Z. 
Wei, M. S. Pshenichnikov, and D. A. Wiersman, Opt. Lett. {\bf 
22}, 102 (1997); M. Nisoli, S. DeSilvestri, O. Svelto, R. Szipocs, 
K. Ferencz, Ch. Spielmann, S. Sartania, and F. Krausz, Opt. Lett. 
{\bf 22}, 522 (1997). 
 
\bibitem{Weiner} A. M. Weiner, Prog. Quant. Electr. {\bf 19}, 161 
(1995); T. Baumert, T. Brixner, V. Seyfried, M. Strehle, G. 
Gerber, Appl. Phys. B, {\bf 65}, 779 (1997). 
 
\bibitem{Warren} C. W. Hillegas, J. X. Tull, D. Goswami, 
D. Strickland, and W. S. Warren, Opt. Lett. {\bf 19}, 737 (1994). 
 
\bibitem{Assion} 
Assion, T. Baumert, M. Bergt, T. Brixner, B. Kiefer, V. Seyfried, 
M. Strehle, and G. Gerber, Science {\bf 282}, 919 (1998) 
 
\bibitem{Gerber} T. Brixner, N. H. Damrauer, and G. Gerber, in Adv. 
At. Mol. Opt. Phys., edited by B. Bederson and H. Walther 
(Academic Press, New York 2001). 
 

 
\bibitem{Murrell} 
W. Murrell in {\it The Bacterial Spore}, edited by G. Gould and A. 
Hurst (Academic Press, 1969). 

\bibitem{Leuschner01}
R.K.G. Leuschner and P.J. Lillford,  
Int. J. Food. Microbiol. {\bf 63,} 35 (2001). 

 
\end{thebibliography}
\end{document}